\documentclass[prb,twocolumn,showpacs]{revtex4}
\usepackage{graphicx}
\usepackage{amsmath}
\begin{document}
\bibliographystyle{apsrev} \title{Annealing mechanisms
  of intrinsic defects in 3C-SiC: a theoretical study}

\author{Michel Bockstedte} 
\author{Alexander Mattausch}
\author{Oleg Pankratov} 
\affiliation{Lehrstuhl f. Theoretische
    Festk\"orperphysik, Universit\"at Erlangen-N\"urnberg, Staudtstr. 7\,B2,
    D-91058 Erlangen, Germany} \date{\today}
\begin{abstract}
  The annealing kinetics of mobile intrinsic defects is investigated by an
  \emph{ab initio} method based on density functional theory. The
  interstitial-vacancy recombination, the diffusion of vacancies and
  interstitials to defect sinks (e.g. surfaces or dislocations) as well as the
  formation of interstitial-clusters are considered. The calculated migration
  and reaction barriers suggest a hierarchical ordering of competing annealing
  mechanisms.  The higher mobility of carbon and silicon interstitials as
  compared to the vacancies drives the annealing mechanims at lower
  temperatures including the vacancy-interstitial recombination and the
  formation of interstitial carbon clusters. These clusters act as a source
  for carbon interstials at elevated temperatures. In p-type
  material we discuss the transformation of the silicon vacancy into the more
  stable vacancy-antisite complex as an annealing mechanism, which is
  activated before the vacancy migration. Recent annealing studies of
  vacancy-related centers in irradiated 3C- and 4H-SiC and semi-insulating 4H-SiC are
  interpreted in terms of the proposed hierarchy of annealing mechanisms.
\end{abstract}
\maketitle
\section{Introduction}
\label{sec:intro}
The unique features of silicon carbide, such as the wide band gap and the
electrical and thermal stability, recommend this semiconductor for high power,
high frequency and high temperature applications. As for any semiconductor,
unwanted defects are introduced into this material by growth processes and
ion-implantation of dopants. Besides the primary defects -- interstitials and
vacancies -- secondary defects such as antisites and defect clusters are
created.  Thermal annealing is applied to reduce these defects to an
inevitable abundance. Thereby the high electron mobility is restored and the
electrical activation of dopants is achieved.

The annealing properties of intrinsic and impurity-related defect centers were
studied in irradiated SiC by various experimental techniques, for instance
electron spin resonance techniques (EPR), deep level transient spectroscopy
(DLTS), photoluminescence spectroscopy (PL) and positron annihilation
spectroscopy (PAS).  Some of the reported centers persist up to a temperature
range of 1300$^{\circ}$C-1700$^{\circ}$C.  Examples for such defects are the
PL-centers D$_{\text{I}}$ (Refs.~\onlinecite{choyke:71,patrick:72a,egilsson:99}) and
D$_{\text{II}}$ (Refs.~\onlinecite{patrick:73,sridhara:98a}) as well as the DLTS-center Z$_{1}$/Z$_{2}$
and E$_{1}$/E$_{2}$ in 4H- and 6H-SiC (Ref.~\onlinecite{kawasuso:01}) respectively.  The
microscopic origin of most of the centers is still investigated. The PL-centers
P$-$U (Ref.~\onlinecite{evans:02}) and D$_{\mathrm{II}}$ have been interpreted as carbon
interstitials\cite{evans:02} or interstitial clusters,\cite{patrick:73} due to
their carbon-related localized vibrational modes (LVMs) with frequencies above
the SiC bulk-spectrum.  Depending on the experimental conditions some of these
centers even grow in intensity during the heat treatment in a temperature
range between 500$^{\circ}$C and 1200$^{\circ}$C, when others (e.g.
vacancy-related centers) gradually
vanish.\cite{freitas:87,egilsson:99a,kawasuso:01}

Vacancies and interstitials act as vehicles for otherwise immobile point
defects. Defect clusters may only be diminished by reacting with or separating
into these mobile entities.  Vacancy related defects have been studied by
PAS\cite{kawasuso:96,brauer:96,kawasuso:98,polity:99,kawasuso:01} and
EPR\cite{itoh:97a,wimbauer:97,vonbardeleben:00,son:01} in irradiated material.
In PAS,\cite{kawasuso:96} defects related to silicon and carbon vacancies can
be distinguished thanks to a considerable difference of the positron
lifetimes, as predicted by theory.\cite{brauer:96a,staab:01} EPR-centers have
also been identified as isolated
silicon\cite{itoh:97a,wimbauer:97,vonbardeleben:00} and
carbon\cite{son:01,konovalov:01,bratus:01} vacancies and the assignment has
been verified
theoretically.\cite{wimbauer:97,petrenko:01,bockstedte:02,bockstedte:03a}
Characteristic differences were observed for the annealing behavior of carbon
and silicon vacancies in irradiated SiC.  A lower annealing temperature
($\sim$500$^{\circ}$C) was deduced for the carbon vacancy by
PAS\cite{kawasuso:96} and EPR.\cite{son:01}  The
annealing of the silicon vacancy occurrs in several stages, at
150$^{\circ}$C, 350$^{\circ}$C and 750$^{\circ}$C (the final annealing stage)
as observed by EPR\cite{itoh:97a} in 3C-SiC in agreement with
PAS\cite{kawasuso:98} data.  In irradiated n-type 4H- and 6H-SiC, however,
additional annealing stages of centers related to a silicon vacancy were
observed by PAS\cite{kawasuso:97,kawasuso:01} at temperatures up to
1450$^{\circ}$C. This finding was explained by the formation of stable
nitrogen-vacancy complexes.  On the other hand, a vacancy-antisite complex was
identified by EPR experiments\cite{lingner:01} in irradiated n-type 6H-SiC
annealed above 750$^{\circ}$C. An EPR-center with similar properties was also
identified in 3C-SiC.\cite{son:97} The center evolves from the silicon vacancy
as a consequence of the vacancy's metastability in p-type and intrinsic
material predicted previously by theory.\cite{rauls:00,bockstedte:00} In both
experiments the signature of the silicon vacancy was absent.  Other
experiments\cite{konovalov:01,son:03} indicate that the carbon and silicon
vacancies posses a higher thermal stability in semi-insulating than in
irradiated SiC.  A comprehensive interpretation of the annealing experiments
in terms of simple mechanisms is lacking sofar. In particular, the role that
interstitials and vacancies play in the reported annealing stages is not
known.  Furthermore, the Fermi-level effect on the annealing behavior of the
charged defects has not been clarified.

\begin{figure}
\includegraphics[width=0.85\linewidth]{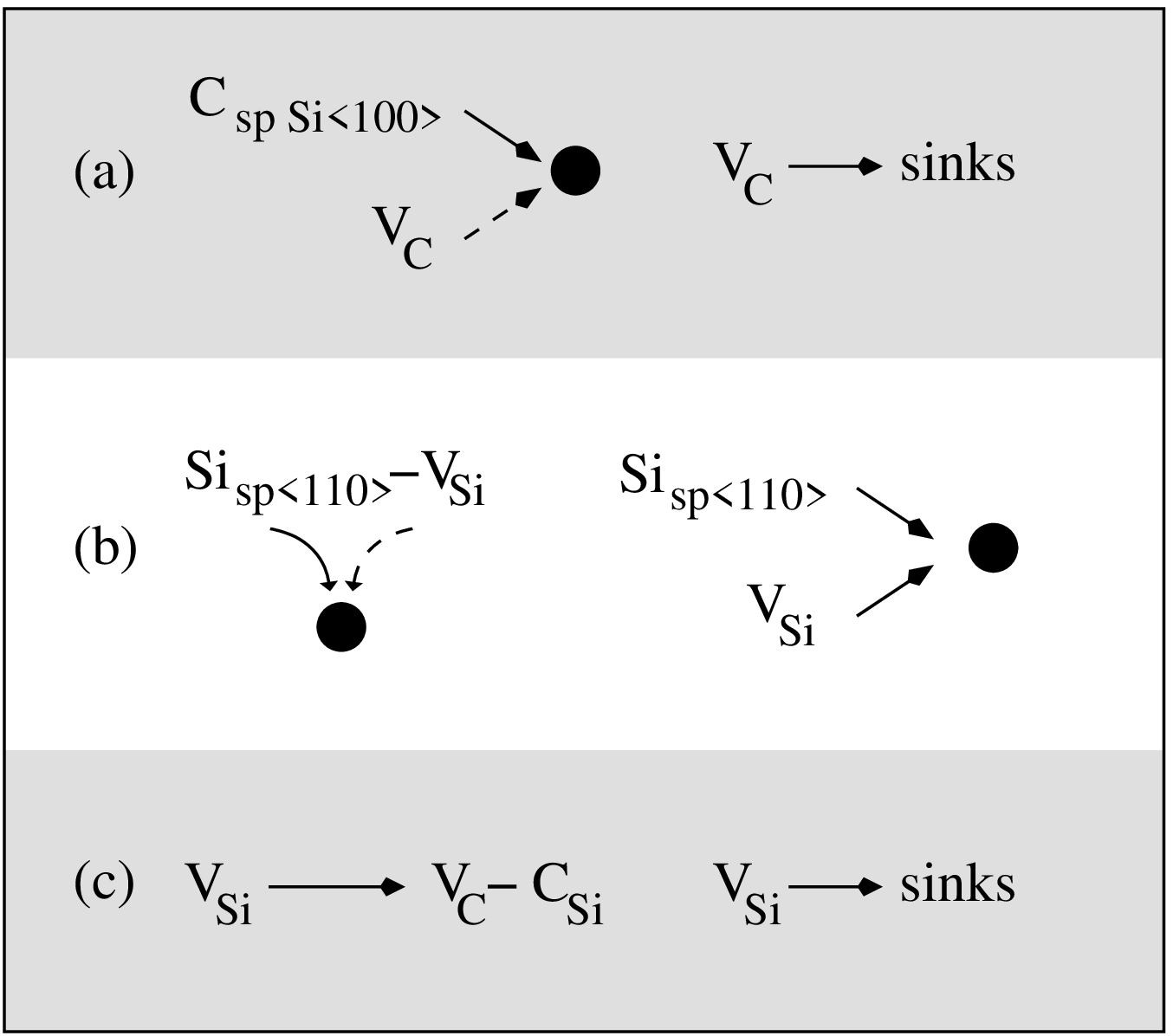}
\caption{\label{fig:scn.anneal} Possible annealing paths of vacancies: 
  (a) carbon vacancy: recombination of a carbon split-interstitial
  (C$_{\mathrm{sp}\langle100\rangle}$) with a carbon vacancy
  (V$_{\mathrm{C}}$) and the migration of carbon vacancies to sinks, (b)
  silicon vacancy: Frenkel pair recombination (compensated and n-type
  material) and diffusion limited recombination with split interstitial
  (Si$_{\mathrm{sp}\langle110\rangle}$), and (c) annealing by the
  metastability induced transformation into a carbon vacancy-antisite complex
  (p-type and compensated material) and the migration of silicon vacancies to
  sinks (n-type). }
\end{figure}

In the present paper we analyze the annealing of vacancies and interstitials
in 3C-SiC at a microscopic level. We study the effect of clustering on
the defect kinetics in case of carbon interstitials.  An \emph{ab initio}
method within the framework of density functional theory (DFT) is employed to
calculate the migration barriers and recombination paths. The
aggregation of carbon interstititials is analysed and the dissociation energy
needed to re-emit a single carbon interstitial is calculated. A hierarchy of
annealing mechanisms (cf. Fig.~\ref{fig:scn.anneal}) is proposed on the basis
of the calculated barriers and dissociation energies at different doping
conditions.  For the carbon-related defects we show that the
vacancy-interstitial recombination and the diffusion-limited annealing of
carbon interstitials are the first two annealing stages in this hierarchy
followed by the diffusion-based annealing of carbon vacancies.  Besides the
diffusion-limited vacancy-interstitial recombination the clustering of carbon
interstitials may facilitate their annealing. Thermally stable carbon clusters
re-emit carbon interstitials at high temperatures thereby contributing to
the kinetics of other thermally stable defects.

While the Fermi-level effect is not pronounced in the case of carbon
vacancies and interstitials, it substantially affects the annealing of silicon
vacancies and interstitials.  This is related to the metastability of the
silicon vacancy in p-type and intrinsic material and the existence of two
different interstitial configurations in p-type and intrinsic or n-type SiC.
For intrinsic (compensated) SiC, the vacancy-interstitial recombination
preceeds the metastability-related transformation of the vacancy into the
carbon vacancy-antisite complex, which in a next stage either dissociates or
anneals via a diffusion-limited mechanism. For p-type conditions the
transformation-induced annealing dominates over the diffusion-limited
vacancy-interstitial recombination. In n-type material a diffusion-limited
annealing of the silicon vacancy should follow the vacancy-interstitial
recombination in the hierarchy of annealing mechanisms.

The outline of the paper is as follows.  Preceeded by a description of the
method in Section~\ref{sec:dft}, we summarize the migration mechanisms of
interstitials and vacancies in Section \ref{sec:v.i.mig}. We also review the
metastability of the silicon vacancy there and analyze the kinetic aspects of
this metastability with emphasis on Fermi-level effects.
Section~\ref{sec:anneal} treats the stable Frenkel pairs and the recombination
paths of vacancies and interstitials.  In Section~\ref{sec:cluster} the
properties of carbon interstitial clusters and their dissoziation energies are
described. The hierarchy of annealing mechanisms is outlined and discussed in
the light of recent experiments in Section~\ref{sec:discuss}.
\section{Method}
\label{sec:dft}
We employ the plane wave pseudopotential program package
\textsf{FHI96MD}~\cite{bockstedte:97a} based on density functional
theory~\cite{hohenberg:64,kohn:65} and the local density
approximation~\cite{perdew:81,ceperley:80} (LDA) for the exchange-correlation
functional. Spin effects are included within the LSDA where noted.  Defects
and their environment are described using large supercells, equivalent to 64
and 216 crystal lattice sites.  A special k-point mesh~\cite{monkhorst:76}
with 8 \textbf{k}-points in the Brillouin zone (2$\times$2$\times$2-mesh) is used. For
the description of Jahn-Teller distortions we performed the calculations in
216 atoms cell at the $\Gamma$-point in order to maintain the 
degeneracy of the defect levels.  The ionization levels of charged defects
were calculated following the approximate procedure proposed by Makov and Payne~\cite{makov:95} to account for the error introduced by the
compensating background.  Especially for the highly charged interstitial
defects this procedure leads to more consistent results.\cite{lento:02,bockstedte:03b}  Norm-conserving
carbon and silicon pseudopotentials of the Troullier-Martins
type~\cite{troullier:91,fuchs:99} are employed. The carbon pseudopotential has
been optimized\cite{pseudo} for calculations with a small basis set.
Extensive tests showed that a basis set of plane-waves with a kinetic
energy up to 30\,Ry yielded practically converged total energy differences.
The coordinates of all atoms in the simulation cell have been relaxed. Test
have shown that all relevant relaxations are well contained within the simulation cells.
\subsection{Formation energies}
The abundance of defects is determined by their effective formation energy $E_{\mathrm{f}}$. For
charged defects, the latter depends on the doping conditions via the Fermi
level $\mu_{\mathrm{F}}$ and on the stoichiometry of the material via the
carbon and silicon chemical potentials $\mu_{\mathrm{C}}$ and
$\mu_{\mathrm{Si}}$. In equilibrium $\mu_{\mathrm{C}}$ and $\mu_{\mathrm{Si}}$
are not independent but related to the chemical potential of SiC
$\mu_{\mathrm{SiC}}$ through
$\mu_{\mathrm{SiC}}=\mu_{\mathrm{C}}+\mu_{\mathrm{Si}}$. Choosing
$\mu_{\mathrm{Si}}$ as a free parameter, we express it as
$\mu_{\mathrm{Si}}=\mu_{\mathrm{Si}}^{0}+\Delta\/\mu$ in terms of the chemical
potential of crystalline Si and the difference $\Delta\/\mu$ to this value.  
The foramtion energy is given by
\begin{equation}
E_{\mathrm{f}}=E_{\mathrm{D,\,cell}}-n_{\mathrm{Si}}\,\mu_{\mathrm{Si}}-n_{\mathrm{C}}\,\mu_{\mathrm{C}}-n_{\mathrm{e}}(E_{\mathrm{V}}+\mu_{\mathrm{F}})\quad,
\end{equation}
where $E_{\mathrm{D,\,cell}}$ is the total energy  of a super cell containing the
defect and $\mu_{\mathrm{F}}$
is defined relative to the valence band edge E$_{\mathrm{V}}$.
As
outlined in Ref.~\onlinecite{bockstedte:03b} E$_{\mathrm{f}}$ may be
conviniently written as
\begin{equation}
E_{{\rm f}}=E_{{\rm D}}^{n_{\rm e}}-(n_{{\rm Si}}-n_{{\rm C}})\,\Delta\mu-n_{{\rm e}}\,\mu_{{\rm F}}\quad.
\end{equation}
In the above equations $n_{\rm e}$ is the number of excess electrons at the
defect, n$_{\mathrm{Si}}$ and $n_{\mathrm{C}}$ are the numbers of silicon and
carbon atoms in the supercell.  The quantity $E_{\mathrm{D}}^{n_{\mathrm{e}}}$
is determined by $E_{{\rm D}}^{n_{\rm e}}=E_{{\rm D,\ cell}}^{n_{\rm
    e}}-n_{{\rm C}}\,\mu_{{\rm SiC}}-(n_{{\rm Si}}-n_{{\rm C}})\,\mu_{{\rm
    Si}}^{0}-n_{\mathrm{e}}\/E_{\mathrm{V}}$ where the chemical potentials
$\mu_{\mathrm{SiC}}$ and $\mu_{\mathrm{Si}}^{0}$ are to a good approximation
given by total energies of the perfect crystals. For $E_{\mathrm{V}}$ we take
the value from a defect-free cell including a correction for the
difference of the average potential in the defect and defect-free cell
(c.f Ref.~\onlinecite{bockstedte:03b}).  To ensure the stability of
SiC, $\Delta\mu$ may only vary between the heat of formation of the
corresponding polytype $H_{{\rm f,\ SiC}}$ and 0, reflecting C-rich and
Si-rich conditions respectively. $H_{\mathrm{f, SiC}}$ only slightly depends
on the polytype and amounts to 0.61\,eV according to our findings choosing
diamond as a reference for $\mu_{\mathrm{C}}$. This is slightly lower than
the experimental value of 0.72\,eV.

The ionization level of charged defects corresponds to the value of $\mu_{\mathrm{F}}$ at
which the defect alters its charge state. Thus it determines the defect
charge state for given doping conditions. The ionization level is obtained from the formation energy by
\begin{equation}
\varepsilon(q_{2}\vert q_{1})=E_{\rm D}^{q_{1}}-E_{\rm D}^{q_{2}}\quad.
\end{equation}
Here $q_{1}$ and $q_{2}$ indicate the different charge states of the defect.
Usually the charge state changes by one electron. However, in some
cases~\cite{baraff:79} it may change by two when the additional
electron-electron repulsion can be compensated by a lattice relaxation. This
effect, known as the negative-U effect, leads to an attractive effective
electron-electron interaction so that the ionization level
$\varepsilon(q+1|q)$ appears below $\varepsilon(q|q-1)$.  The realization of a
charge state as a ground state for implies that the corresponding ionization
levels and the occupied single particle levels are located below the
conduction band edge. Even though a physical relevance of the Kohn-Sham levels
is not rigorously founded by DFT, they reproduce the experimental
quasi-particle band structure quite well. Yet, the calculated Kohn-Sham band
gaps (1.2\,eV and 2.2\,eV in 3C- and 4H-SiC) are typically smaller than the
experimental values (2.39\,eV and 3.27\,eV). In order to assess the defect
charge states we follow a common practice and use the experimental value for
the conduction band edge. However this procedure cannot unambigiously describe the
charge states that are stable only in the vicinity of the
conduction band edge.

\subsection{Defect migration}
The analysis of migration and recombination paths starts with establishing  the stable initial
and final configuration. Then standard search methods are used to find the path
between these configurations and to determine the saddle point.  For the interstitial
migration and the Frenkel-pair recombination we employ an
implementation~\cite{pehlke:00p} of the ridge method of Ionova and
Carter.\cite{ionova:93} The ridge method performs an automatic search for the
saddle point starting from the initial and final configuration. We repeat the
search for all relevant charge states and analyze the obtained saddle points
with respect to the geometry and the electronic structure.  When the 
migration path depends on the occupation of defect states (near degeneracies,
Jahn-Teller distorted ground states) this automatic search fails.
This is relevant for the vacancy migration and the metastability-induced
transformation of the silicon vacancy into a vacancy-antisite complex as
discussed in Ref.~\onlinecite{bockstedte:03b}. In
these cases, we analyze the potential energy surfaces using the drag method.
The procedure is outlined in Ref.~\onlinecite{bockstedte:03b}, where we
discuss the migration of the vacancies and their transition states in detail.
\section{Migration of interstitials and vacancies}
\label{sec:v.i.mig}
\subsection{\label{sec:i}interstitials}
In 3C-SiC a large number of interstitial configurations exists. Besides the
tetrahedral and hexagonal interstitials also split-interstitials are relevant.
In the latter the interstitial atom and the lattice atom share a lattice
site in a dumbbell-like configuration, which is preferentially oriented along
the $\langle100\rangle$- or $\langle110\rangle$-direction.  Recently, we have
analyzed these configurations for silicon and carbon
interstitials.\cite{mattausch:01,bockstedte:03b} In particular, we
have discerned the energetically most favorable sites. At these sites the
interstitial migration starts and ends. Other interstitial configurations act
as intermediate states along the migration path.  The migration path and the
energy barriers were investigated using a 64-atom
cell and special k-points ($2\times2\times2$ Monkhorst-Pack mesh).
\begin{table}
\caption{\label{tab:ion}Ionization levels of the mobile defects in
  3C-SiC given relative to the valence band edge .}
\begin{ruledtabular}
\begin{tabular}{cccccc}
&$\varepsilon(1^{+}|2^{+})$&$\varepsilon(0|1^{+})$&$\varepsilon(1^{-}|0)$&$\varepsilon(2^{-}|1^{-})$\\
\hline
V$_{\mathrm{C}}$&1.29&1.14&2.69&2.04\\
V$_{\mathrm{Si}}$&\multicolumn{1}{c}{--}&0.18&0.61&1.76\\
V$_{\mathrm{C}}$-C$_{\mathrm{Si}}$&1.24&1.79&2.19&\multicolumn{1}{c}{--}\\
\hline
Si$_{\mathrm{sp}\langle110\rangle}$&0.4&1.1&\multicolumn{1}{c}{--}&\multicolumn{1}{c}{--}\\
C$_{\mathrm{sp}\langle100\rangle}$&0.6&0.8&1.8&\multicolumn{1}{c}{--}\\
C$_{\mathrm{sp\/Si}\langle100\rangle}$&0.4&0.7&1.9&2.3
\end{tabular}
\end{ruledtabular}
\end{table}

\begin{figure*}
\includegraphics[width=\linewidth]{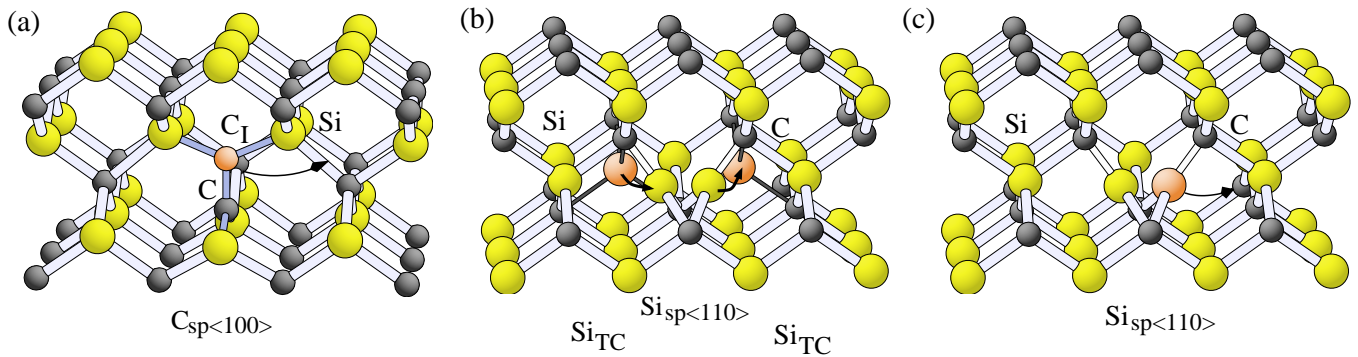}
\caption{\label{fig:mig.i}Migration of interstitial: (a) migration of the
  carbon split-interstitial by second neighbor hops, (b) kick-out migration
  of the silicon interstitial between adjacent
  Si$_{\mathrm{T\/C}}$-sites via the split-interstitial
  Si$_{\mathrm{sp}\langle110\rangle}$, and (c) migration of the silicon interstitial
  between adjacent Si$_{\mathrm{sp}\langle110}$ configurations.}
\end{figure*}

Among the carbon interstitials the two split-interstitials 
C$_{\mathrm{sp}\langle100\rangle}$ and C$_{\mathrm{sp\/Si}\langle100\rangle}$
are the most favorable configurations. The split-interstitial C$_{\mathrm{sp}\langle100\rangle}$
is centered at a carbon site and oriented in the $\langle100\rangle$-direction
as displayed in Fig.~\ref{fig:mig.i}a. It is more favorable than the
C$_{\mathrm{sp\/Si}\langle100\rangle}$-interstitial which is a $\langle100\rangle$-oriented silicon-carbon
dumbbell at a silicon site. The electronic structure of the interstitials is
discussed in detail in Ref.~\onlinecite{bockstedte:03b}. The two
interstitials are stable in positive and negative charge states. The ionization
levels of C$_{\mathrm{sp}\langle100\rangle}$ and
C$_{\mathrm{sp\/Si}\langle100\rangle}$ are given in Tab.~\ref{tab:ion}. Other
interstitial sites, such as the hexagonal site and the tetrahedral sites, are
higher in energy.

Two alternative migration paths have been found for the migration of carbon
interstitials: a migration by second neighbor hops of
C$_{\mathrm{sp}\langle100\rangle}$ between adjacent carbon-sites and a
migration by nearest neighbor hops between adjacent carbon and silicon lattice
sites, going through the sequence
C$_{\mathrm{sp}\langle100\rangle}\rightarrow$C$_{\mathrm{sp\/Si}\langle100\rangle}\rightarrow$C$_{\mathrm{sp}\langle100\rangle}$.
The migration by second neighbor hops is indicated in Fig.~\ref{fig:mig.i}.

The migration barriers for the two mechanisms are given in
Tab.~\ref{tab:c.mig}. The nearest neighbor hop
C$_{\mathrm{sp}\langle100\rangle}\rightarrow$
C$_{\mathrm{sp\/Si}\langle100\rangle}$ and the second neighbor hop
C$_{\mathrm{sp}\langle100\rangle}\rightarrow$C$_{\mathrm{sp}\langle100\rangle}$
have similar migration barriers, except for the
C$_{\mathrm{sp}\langle100\rangle}^{2+}$.  Small barriers are found for the
migration of the positive, neutral and negative interstitials.
The analysis of migration paths via other intermediate interstitial sites
indicates that such paths are unlikely: the energy differences between
C$_{\mathrm{sp}\langle100\rangle}$ and these sites typically exceed the small migration
barriers found for the above mechanisms.

\begin{table}
\caption{\label{tab:c.mig} Migration of carbon interstitials: energy barriers
 of nearest neighbor hops
  and second neighbor hops for the relevant charge states. The migration
  of C$_{\mathrm{sp}\langle100\rangle}^{0}$ starts from the tilted configuration.}
\begin{ruledtabular}
\begin{tabular}{lccccccc}
\multicolumn{1}{c}{path}&\multicolumn{5}{c}{energy barrier [\,eV\,]}\\
\hline
& 2$^{+}$&&1$^{+}$&&0&&1$^{-}$\\
\cline{2-2}\cline{4-4}\cline{6-6}\cline{8-8}
C$_{\mathrm{sp}\langle100\rangle}\rightarrow$
C$_{\mathrm{sp\/Si}\langle100\rangle}$&1.7&&0.9&&0.5&&0.7\\
C$_{\mathrm{sp\/Si}\langle100\rangle}\rightarrow$
C$_{\mathrm{sp}\langle100\rangle}$&0.9&&0.2&&0.2&&0.1\\[1ex]
C$_{\mathrm{sp}\langle100\rangle}\leftrightarrow$
C$_{\mathrm{sp}\langle100\rangle}$&1.4&&0.9&&0.5&&0.6\\
\end{tabular} 
\end{ruledtabular}
\end{table}

Among the silicon interstitials the important sites are the tetrahedrally
carbon coordinated site (Si$_{\mathrm{T\/C}}$) and the (110)-oriented
split-interstitial (Si$_{\mathrm{sp}\langle110\rangle}$). The tetrahedral
interstitial dominates in p-type material. Our analysis of the electronic
structure shows~\cite{bockstedte:03b} that it has no deep defect states within
the band gap and therefore has a charge state of 4$^{+}$.  The bare silicon
ion is efficiently screened by a crystal polarization. Considering the
possible occupation of a defect state located slightly above the experimental
conduction band edge we concluded in Ref~\onlinecite{bockstedte:03b} that the
charge state 4$^{+}$ prevails for Fermi-levels below the mid-gap (by the
large Madelung corrections $\varepsilon(3^{+}|4^{+})$ is placed above
the mid-gap). This is also consistent with our recent findings for
Si$_{\mathrm{T\/C}}$ in 4H-SiC.  The split-interstitial
Si$_{\mathrm{sp}\langle110\rangle}$ dominates in intrinsic and n-type
material. It has several deep states within the band gap and is stable in the
charge states 2$^{+}$, 1$^{+}$, and the neutral state.  The ionization levels
are given in Tab.~\ref{tab:ion}. For the diffusion in intrinsic and n-type
material the neutral configuration is most important.

In p-type material all migration paths begin and end at the
Si$_{\mathrm{T\/C}}$ site. The migration of Si$_{\mathrm{T\/C}}$ proceeds
basically by two different mechanisms: (i) a kick-out mechanism via via the
split-interstitials Si$_{\mathrm{sp}\,\langle100\rangle}$ and
Si$_{\mathrm{sp}\langle110\rangle}$ as intermediate sites and (ii) a direct
migration via a silicon coordinated interstitial site Si$_{\mathrm{T\/Si}}$.
In the kick-out mechanism the tetrahedral interstitial moves towards one of
its silicon neighbors and forms a split-interstitial as an intermediate site
as indicated in Fig.~\ref{fig:mig.i}.
To complete the migration, one of the silicon atoms of the split-interstitial
moves to a neighboring Si$_{\mathrm{T\/C}}$-site.  The energy barriers for the
migration via the different intermediate sites are listed in
Tab.~\ref{tab:si.mig}.  The lowest barrier is obtained for the migration via
Si$_{\mathrm{sp}\langle110\rangle}$.  The barriers for the other paths are
only slightly higher 0.1\,eV.  In case of the kick-out mechanism the
transition state lies in the vicinity of the intermediate site, which is found
to be hardly stable.  For the direct path the intermediate
Si$_\mathrm{{T\/Si}}$ is more stable with a barrier of 0.3\,eV.

\begin{table}
\caption{\label{tab:si.mig} Migration of silicon interstitials:
  energy barriers $E_{\mathrm{m}}$ in eV for different migration paths. The
  dominant migration paths are shown in Fig.~\ref{fig:mig.i}. The charge state is
  indicated. See text
  for details.}
\begin{ruledtabular}
\begin{tabular}{cclclc}
&Si$_{\mathrm{T\/C}}^{4+}$$\leftrightarrow$Si$_{\textrm{sp}\langle100\rangle}^{4+}$
&&Si$_{\mathrm{T\/C}}^{4+}$$\leftrightarrow$Si$_{\mathrm{sp\langle110\rangle}}^{4+}$
&&Si$_{\mathrm{T\/C}}^{4+}\leftrightarrow$Si$_{\mathrm{T\/Si}}^{4+}$ \\
\cline{2-2}\cline{4-4}\cline{6-6}
1$\rightarrow$2&3.5&&3.4&&3.5\\
2$\rightarrow$1&0.03&&0.05&&0.3\\[2ex]
&Si$_{\mathrm{sp}\langle110\rangle}^{+}$&&Si$_{\mathrm{sp}\langle110\rangle}^{0}$
&&Si$_{\mathrm{sp}\langle110\rangle}^{0}$\\
&$\leftrightarrow$\,Si$_{\mathrm{sp}\langle110\rangle}^{+}$&&$\leftrightarrow$\,Si$_{\mathrm{sp}\langle110\rangle}^{0}$&&$\leftrightarrow$\,Si$_{\mathrm{sp}\langle1\bar{1}0\rangle}^{0}$\\
\cline{2-2}\cline{4-4}\cline{6-6}
1$\leftrightarrow$2&1.0&&1.4&&1.3\\[2ex]
\end{tabular} 
\end{ruledtabular}
\end{table}
In intrinsic (compensated) and n-type material the interstitial migration of
Si$_{\mathrm{sp}\langle110\rangle}^{0}$ dominates. As indicated in
Fig.~\ref{fig:mig.i}c the migration proceeds by hops of one of the silicon
atoms to the next silicon lattice site, where a new
Si$_{\mathrm{sp}\langle110\rangle}$-interstitial is formed.  A barrier of
1.4\,eV is found for this hop. The interstitial may also change its
orientation. For example, by the rotation of the pair around the (001)-axis
the orientation changes from (110) to $(1\bar{1}0)$. This process is
associated with a barrier of 1.3\,eV.  We have also considered the positive
interstitial, which migrates with a barrier of 1.0\,eV.

So far the migration of interstitials has not been investigated on an \emph{ab
  initio} level in 4H-SiC. However, our analysis of the ground state
configurations shows\cite{bockstedte:03} that the carbon split-interstitials
at the carbon and silicon lattice sites have identical properties as in
3C-SiC, except that C$_{\mathrm{sp}\langle100\rangle}^{2-}$ becomes stable in
n-type 4H-SiC. For the silicon interstitials in 4H-SiC, besides the sites
discussed for 3C-SiC, additional sites appear: a site where the interstitial
has simultaneously four carbon and silicon neighbors, and an open cage formed
by the hexagonal rings of two adjacent lattice planes. For the interstitial
sites with cubic environment we have found similar properties to 3C-SiC, also
with respect to the (110)-oriented split-interstitial.  In the open cage a
position close to three carbon neighbors is preferred.  The interstitial with
four silicon and carbon neighbors is energetically unfavorable and may
represent an intermediate site of a migration path. From the bonding
properties of the interstitials we expect similar migration barriers as in
3C-SiC.
\subsection{\label{sec:v}Vacancy migration}
The ground state properties of the carbon and silicon vacancy have been
discussed in detail by Zywietz~\emph{et~al}.~\cite{zywietz:99} and Torpo
\emph{et~al}.~\cite{torpo:01}. Here we briefly summarize a few facts for the sake
of clarity. Both vacancies have a threefold degenerate level within the band
gap. Its occupation leads in case of the carbon vacancy (V$_{\mathrm{C}}$) to
a considerable Jahn-Teller-relaxation. The vacancy can exist in
the charge states 2$^{+}$, $1^{+}$, and neutral. Our calculation indicate that
the negative charge states should not be realized. The ionization levels are
listed in Tab.~\ref{tab:ion}. According to our calculations the positive charge
state is unstable due to a negative-U behavior. However, for $U$,
which is given by $U=\varepsilon(+|++)-\varepsilon(0|+)$, we obtain a rather
small value of $-0.15$\,eV. This value should be considered as an estimate that depends on the size
of the supercell and on the correction for the total energy of the charge defect.\cite{makov:95}
Neglecting this correction, Zywietz \emph{et~al}.\cite{zywietz:99} arrived
at $U=-0.48$\,eV. In contrast to our findings, the calculations of Torpo
\emph{et al}.\cite{torpo:01} suggest the absence of a negative-U behavior.
However, the deviations in the results for $U$ are within the uncertainty in
the evaluation of total energies for charged defects of about 0.2\,eV.
Irrespective whether a small negative-U is present or not, the positive charge
state becomes relevant for $\mu_{\mathrm{F}}$ around the ionization level
$\varepsilon(0|++)=1.2$\,eV, in particular, when the Fermi level is pinned by
an excess concentration of vacancies. This is also suggested by the recent identification of the
EI5-center in 4H-SiC as a positive carbon vacancy,\cite{son:01,bockstedte:02}
which shows a similar behavior as V$_{\mathrm{C}}^{+}$ in 3C-SiC.

The DFT-LDA calculations predict that the silicon vacancy prefers a high-spin
state over a Jahn-Teller-distorted ground state. This is consistent with the
findings of spin-resonance experiments\cite{itoh:97a,wimbauer:97} for
V$_{\mathrm{Si}}^{-}$. According to our analysis the silicon vacancy is
realized in the charge states 1$^{+}$, neutral, 1$^{-}$, and 2$^{-}$ in
3C-SiC.  Already in p-type material the vacancy becomes neutral and changes to
a negative charge state for a Fermi-level at 0.6\,eV. The charge state 2$^{-}$
is stabilized in n-type material.

In 4H-SiC the carbon and silicon vacancy possess similar properties.  The
occupation of the lattice sites with local cubic and hexagonal coordination
introduces negligible changes for the silicon vacancy. In case of
V$_{\mathrm{Si}}^{-}$ the calculation of HF-tensors\cite{bockstedte:03a} and
EPR-experiments\cite{wimbauer:97} show that the two sites cannot be
distinguished. Also for the carbon vacancy similar formation energies and
ionization levels are obtained at the cubic and hexagonal site. However, the
different rehybridisation of the silicon dangling bonds\cite{bockstedte:03a}
affects the local symmetry and for V$_{\mathrm{C}}^{+}$ leads to different
HF-signatures at the cubic and hexagonal sites.\cite{bockstedte:03a} A
relevant effect for the diffusion in n-type 4H-SiC is the stabilization of
additional negative charge states (V$_{\mathrm{C}}^{-}$, V$_{\mathrm{C}}^{2-}$,
V$_{\mathrm{Si}}^{3-}$, and V$_{\mathrm{Si}}^{4-}$) due to the larger band gap.

\begin{figure}
\includegraphics[width=0.65\linewidth]{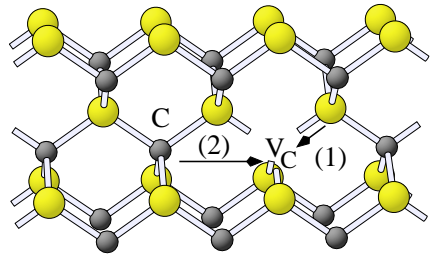}
\caption{\label{fig:mig.v} Migration of the carbon vacancy:
  (1) nearest neighbor hop (2) migration on the carbon sublattice by
      second neighbor hops.}
\end{figure}

For the migration of the vacancies we have analyzed mechanisms that
involve either nearest neighbor hops or second neighbor hops. In
Fig.~\ref{fig:mig.v} both mechanisms are
indicated for the carbon vacancy.  Our analysis has shown\cite{bockstedte:02}
that the silicon and carbon vacancies migrate by second neighbor
hops.\cite{mattausch:01,bockstedte:03b} The arguments that rule out a
migration entirely based on nearest neighbor hops are as follows.  By a
nearest neighbor hop the vacancy transforms into a vacancy-antisite-complex as
the silicon (carbon) neighbor moves onto the sublattice of the opposite kind,
thereby creating an antisite next to the vacant lattice site.  In the case of
the carbon vacancy we have found that the silicon vacancy-antisite-complex is
unstable in all relevant charge states.\cite{bockstedte:00} A barrier that
could stabilize the vacancy-antisite complex is not present.  This finding
rules out any migration mechanism based on nearest neighbor hops for the
carbon vacancy. In case of the silicon vacancy, it has been found by
us\cite{bockstedte:00} in 3C-SiC and by Rauls \emph{et~al}.\cite{rauls:00} for
the neutral complex in 4H-SiC that the carbon vacancy-antisite-complex may be
more stable than the vacancy. This metastability of the silicon vacancy occurs
in p-type and intrinsic (compensated) material and has implications on the
migration as well as the annealing of V$_{\mathrm{Si}}$. It will be considered
in detail in Sec.~\ref{sec:meta}.  However, a migration mechanism entirely
based on nearest neighbor hops is impossible: a consecutive nearest neighbor
hop would transform the V$_{\mathrm{C}}$-C$_{\mathrm{Si}}$-complex into a
V$_{\mathrm{Si}}$-Si$_{\mathrm{C}}$-C$_{\mathrm{Si}}$-complex. We have found
that this complex is unstable.

Our analysis of the second neighbor hop, its migration path and the
calculation of the migration barriers is discussed in detail in
Ref.~\onlinecite{bockstedte:03b}. Here we basically summarize our findings for
the migration barriers as tabulated in Tab~\ref{tab:v.mig}. The barriers are
obtained using the 216 atom cell and the $\Gamma$-point for the k-summation.
This also allowed for the proper inclusion of Jahn-Teller distortions in case
of the carbon vacancy and the inclusion of spin polarization for the silicon
vacancy.  Using a 64 atom cell and special k-points ($2\times2\times2$
Monkhorst-Pack mesh) the calculated barriers agree within 0.3\,eV.  For the
carbon vacancy we find relatively high migration barriers.  They strongly
depend on the charge state and vary between 5\,eV in p-type material
(V$_{\mathrm{C}}^{2+}$) and 3.5\,eV in n-type material (V$_{\mathrm{C}}^{0}$).
The charge state dependence has its origin in the successive occupation of
bonding defect levels at the transition state.  For the silicon vacancy we
obtain smaller migration barriers and a less pronounced dependence on the
charge state. The migration barrier in p-type and intrinsic material with
values between 3.2\,eV and 3.6\,eV is larger than the barrier for the
transformation into the carbon vacancy-antisite-complex. The implications of
this finding are discussed in the next section. In n-type material the migration
barrier amounts to 2.4\,eV (V$_{\mathrm{Si}}^{2-}$). Since the vacancy is
stable in n-type material, a possible transformation of the vacancy here does
not interfere with its migration.

\begin{table}
\caption{\label{tab:v.mig}Vacancy migration 
      and transformation of V$_{\mathrm{Si}}$ into
      V$_{\mathrm{C}}$-C$_{\mathrm{Si}}$: Energy barriers for the relevant
      charge states.}
\begin{ruledtabular}
\begin{tabular}{cclclclclc}
path&\multicolumn{9}{c}{energy barriers [\,eV\,]}\\
\hline
&2$^{+}$&&1$^{+}$&&0&&1$^{-}$&&2$^{-}$\\
\cline{2-2}\cline{4-4}\cline{6-6}\cline{8-8}\cline{10-10}
V$_{\mathrm{C}}$$\rightarrow$V$_{\mathrm{C}}$&5.2&&4.1&&3.5&&-&&-\\
V$_{\mathrm{Si}}$$\rightarrow$V$_{\mathrm{Si}}$&-&&3.6&&3.4&&3.2&&2.4\\[2ex]
V$_{\mathrm{Si}}$$\rightarrow$V$_{\textrm{C}}$-C$_{\textrm{Si}}$&-&&1.9&&2.4 (2.2)&&2.5&&2.7\\
V$_{\textrm{C}}$-C$_{\textrm{Si}}$$\rightarrow$V$_{\mathrm{Si}}$&6.1&&4.2&&3.5 (2.4)&&2.4&&-\\
\end{tabular}
\end{ruledtabular}
\end{table}

\subsection{\label{sec:meta}Metastability of the silicon vacancy}
\begin{figure}
\includegraphics[width=\linewidth]{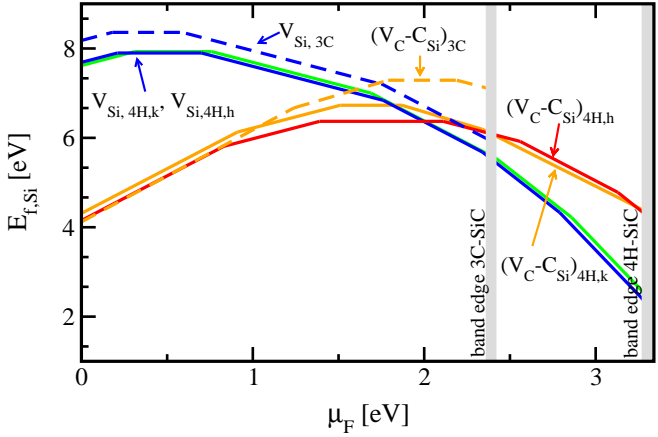}
\caption{\label{fig:vsi-vccsi}Formation energy of the silicon vacancy V$_{\mathrm{Si}}$
  and the carbon vacancy-antisite-complex V$_{\mathrm{C}}$-C$_{\mathrm{Si}}$
  for 3C-SiC and 4H-SiC. The experimental band edges of 3C- and 4H-SiC are indicated by
  vertical bars. The subscripts refer to the polytype and  the
  cubic and hexagonal lattice site (k and h respectively). (V$_{\mathrm{C}}$-C$_{\mathrm{Si}}$)$_{4H,k}$ and
  (V$_{\mathrm{C}}$-C$_{\mathrm{Si}}$)$_{4H,h}$ refer to complexes with the
  vacancy and antisite at either cubic or hexagonal sites that are aligned
  along the c-axis.}
\end{figure}

Our results in 3C- and 4H-SiC for the metastability of the silicon vacancy (in
4H-SiC we have
considered complexes with the vacancy-antisite pair located at neighboring
cubic or hexagonal sites and aligned along the c-axis) and the findings of
Lingner\emph{et~al}.~\cite{lingner:01} suggest that the following discussion
is not restricted to a certain polytype.  The silicon vacancy and the carbon
vacancy-antisite-complex may be transformed into each another by a nearest
neighbor hop as discussed above.  Depending on the doping conditions a
metastability of the silicon vacancy arises, when the
V$_{\mathrm{C}}$-C$_{\mathrm{Si}}$-complex is more stable than the silicon
vacancy. This is the case in p-type and intrinsic (compensated) material. A
stable silicon vacancy is retained for a
Fermi-level position above 1.7\,eV. This is demonstrated in Fig.~\ref{fig:vsi-vccsi}. For a
given Fermi-level, the two defects V$_{\mathrm{Si}}$ and
V$_{\mathrm{C}}$-C$_{\mathrm{Si}}$ possess different charge states, as the
transformation of V$_{\mathrm{Si}}$ into V$_{\mathrm{C}}$-C$_{\mathrm{Si}}$
rises the position of localized states within the band gap.

The energy barriers we have obtained for the transformation
\protect{V$_{\mathrm{Si}}\rightarrow$(V$_{\mathrm{C}}$-C$_{\mathrm{Si}}$)} and
the reverse process
\protect{(V$_{\mathrm{C}}$-C$_{\mathrm{Si}}$)$\rightarrow$V$_{\mathrm{Si}}$}
are listed in Tab.~\ref{tab:v.mig}.  The values refer to our recent results
obtained for 3C-SiC using a 216 atom cell and the $\Gamma$-point and include
spin effects.\cite{bockstedte:03b} The barriers are calculated for each given
charge state. We assume that the initial charge state is preserved during the
transformation and equilibrates at the final configuration.  This applies also
to the transformation of
V$_{\mathrm{Si}}^{2-}\,\rightarrow\,$(V$_{\mathrm{C}}$-C$_{\mathrm{Si}}$)$^{2-}$
and
(V$_{\mathrm{C}}$-C$_{\mathrm{Si}}$)$^{2+}\,\rightarrow\,$V$_{\mathrm{Si}}^{2+}$,
where the final configurations (V$_{\mathrm{C}}$-C$_{\mathrm{Si}}$)$^{-}$ and
V$_{\mathrm{Si}}^{+}$ are the ground state configurations.  For all
charge states the transformation occurs in a low-spin state, except for
V$_{\mathrm{Si}}^{0}$ where the high-spin state has a lower barrier (given
in braces in Tab.~\ref{tab:v.mig}).  Since this high-spin state is not the
ground state of the final V$_{\mathrm{C}}$-C$_{\mathrm{Si}}$-complex, the
complex relaxes into the low-spin state before a reverse transformation
occurs.

The metastability of the vacancy in p-type and its stabilization in n-type
material is accompanied with a pronounced variation of the transformation
barriers with the charge state.  In p-type material the transformation
V$_{\mathrm{Si}}\,\rightarrow\,$V$_{\mathrm{C}}$-C$_{\mathrm{Si}}$ has a
much lower barrier with values between 1.9 and 2.5\,eV than the reverse
transformation with a barrier between 6.1\,eV and 3.5\,eV. The transformation
is thus more likely than the reverse transformation. Only when
(V$_{\mathrm{C}}$-C$_{\mathrm{Si}}$)$^{-}$ becomes stable, i.e. in n-type 3C-SiC,
the transformation of V$_{\mathrm{Si}}^{2-}$ has a lower probability than the
transformation of (V$_{\mathrm{C}}$-C$_{\mathrm{Si}}$)$^{-}$. 

A comparison of the transformation barrier with the migration barrier shows
that the transformation occurs with a much higher probability than the
migration in the charge states 1$^{+}$, neutral and 1$^{-}$, which are relevant
in p-type and compensated material. The reverse transformation
\protect{V$_{\mathrm{C}}$-C$_{\mathrm{Si}}$$\rightarrow$V$_{\mathrm{Si}}$}
has much larger barriers than the migration of V$_{\mathrm{Si}}$. Note that
the relevant charge states are 2$^{+}$ and 1$^{+}$ for this process in p-type
material.  Therefore, the vacancy migration is kinetically hindered for low
and moderate temperatures unless the reverse
transformation is activated at high temperatures.  Then a migration of
V$_{\mathrm{C}}$-C$_{\mathrm{Si}}$ may take place via V$_{\mathrm{Si}}$. The
effective barrier for the migration path V$_{\mathrm{C}}$-C$_{\mathrm{Si}}$$\rightarrow$V$_{\mathrm{Si}}\rightarrow$V$_{\mathrm{Si}}\rightarrow$(V$_{\mathrm{C}}$-C$_{\mathrm{Si}}$) is given by the formation energies
$E_{\mathrm{f}}^{\mathrm{V}_{\mathrm{Si}}}$, its migration barrier
$E_{\mathrm{m}}^{\mathrm{V}_{\mathrm{Si}}}$, and the formation energy
$E_{\mathrm{f}}^{{\mathrm{V}_{\mathrm{C}}}\mathrm{-C}_{\mathrm{Si}}}$ of the
complex
\begin{equation}
E_{\mathrm{m, eff}}=E_{\mathrm{f}}^{\mathrm{V}_{\mathrm{Si}}}-E_{\mathrm{f}}^{{\mathrm{V}_{\mathrm{C}}}\mathrm{-C}_{\mathrm{Si}}}+E_{\mathrm{m}}^{\mathrm{V}_{\mathrm{Si}}}
\end{equation}
In p-type material this effective barrier amounts to 7.7\,eV and drops to a
value of 4.3\,eV for $\mu_{\mathrm{F}}=1.2$\,eV around mid-gap, when the
vacancy is negative. The effective barrier is still 3.2\,eV at
$\mu_{\mathrm{F}}=1.76\,eV$, when the silicon vacancy becomes doubly negative.
Here we have assumed that the charge state equilibrates at the vacancy. If
this is not the case, the effective migration barrier for the positive
and neutral
V$_{\mathrm{C}}$-C$_{\mathrm{Si}}$-complexes are 6.3\,eV and 4.5\,eV respectively.

Besides a migration of a metastable silicon vacancy, also a dissociation of
the carbon vacancy-antisite-complex has to be considered.\cite{mattausch:01}
In this case the carbon vacancy migrates away from the antisite at the expense
of the binding energy. Using the migration barrier for V$_{\mathrm{C}}$
between 3.5\,eV (V$_{\mathrm{C}}^{0}$) to 5\,eV (V$_{\mathrm{C}}^{2+}$) plus
the binding energy of about 1\,eV as an estimate for the dissociation barrier,
we obtain values of 6.2\,eV for (V$_{\mathrm{C}}$-C$_{\mathrm{Si}}$)$^{2+}$
and 4.5\,eV for (V$_{\mathrm{C}}$-C$_{\mathrm{Si}}$)$^{0}$. Only in p-type
material this process has a similar probability as the reverse transformation
followed by a migration of the silicon vacancy.

We have not investigated the transformation barriers in 4H-SiC. Rauls
\emph{et~al}.~\cite{rauls:00} have obtained a transformation barrier of 1.7\,eV
for the neutral vacancy using a DFT-based tight-binding scheme, which is in
agreement with our results. Therefore we expect a similar charge state
dependence of the barriers in 4H-SiC as discussed above. For
the charge states 3$^{-}$ and 4$^{-}$ of the vacancy we expect larger barriers
for the transformation
V$_{\mathrm{Si}}\,\rightarrow\,$(V$_{\mathrm{C}}$-C$_{\mathrm{Si}}$) and
correspondingly lower barriers for the reverse transformation.

\section{Recombination of Frenkel pairs}
\label{sec:anneal}
\subsection{Carbon vacancies}
In a carbon Frenkel pair the interstitial and vacancy may be nearest neighbors or
second neighbors. Also vacancies and
interstitials may attract each other over larger distances. For their
recombination the interstitial has to migrate to the vacancy.
In the following we analyze the recombination of Frenkel pairs where a carbon
split-interstitial is a nearest or second neighbor of the carbon vacancy.
\begin{table}
\caption{\label{tab:i-v-pairs}Recombination of vacancies and interstitials: Energy barriers of
  Frenkel pairs for the relevant charge states. The geometry of the
  Frenkel pairs is displayed in Fig.\ref{fig:i-v-pairs}.}
\begin{ruledtabular}
\begin{tabular}{cclclc}
Frenkel pair&\multicolumn{5}{c}{energy barriers [\,eV\,]}\\
\hline
&$2^{+}$&&$1^{+}$&&0\\
\cline{2-2}\cline{4-4}\cline{6-6}
C$_{\mathrm{sp}\langle100\rangle}$-V$_{\mathrm{C}}$&1.0&&0.5&&0.4\\
C$_{\mathrm{sp\/Si}\langle100\rangle}$-V$_{\mathrm{C}}$&1.2&&1.2&&1.4\\[1ex]
Si$_{\mathrm{sp}\langle110\rangle}$-V$_{\mathrm{Si}}$&0.0&&$\sim0.2$&&0.2\\
\end{tabular}
\end{ruledtabular}
\end{table}
\begin{figure}
\includegraphics[width=0.9\linewidth]{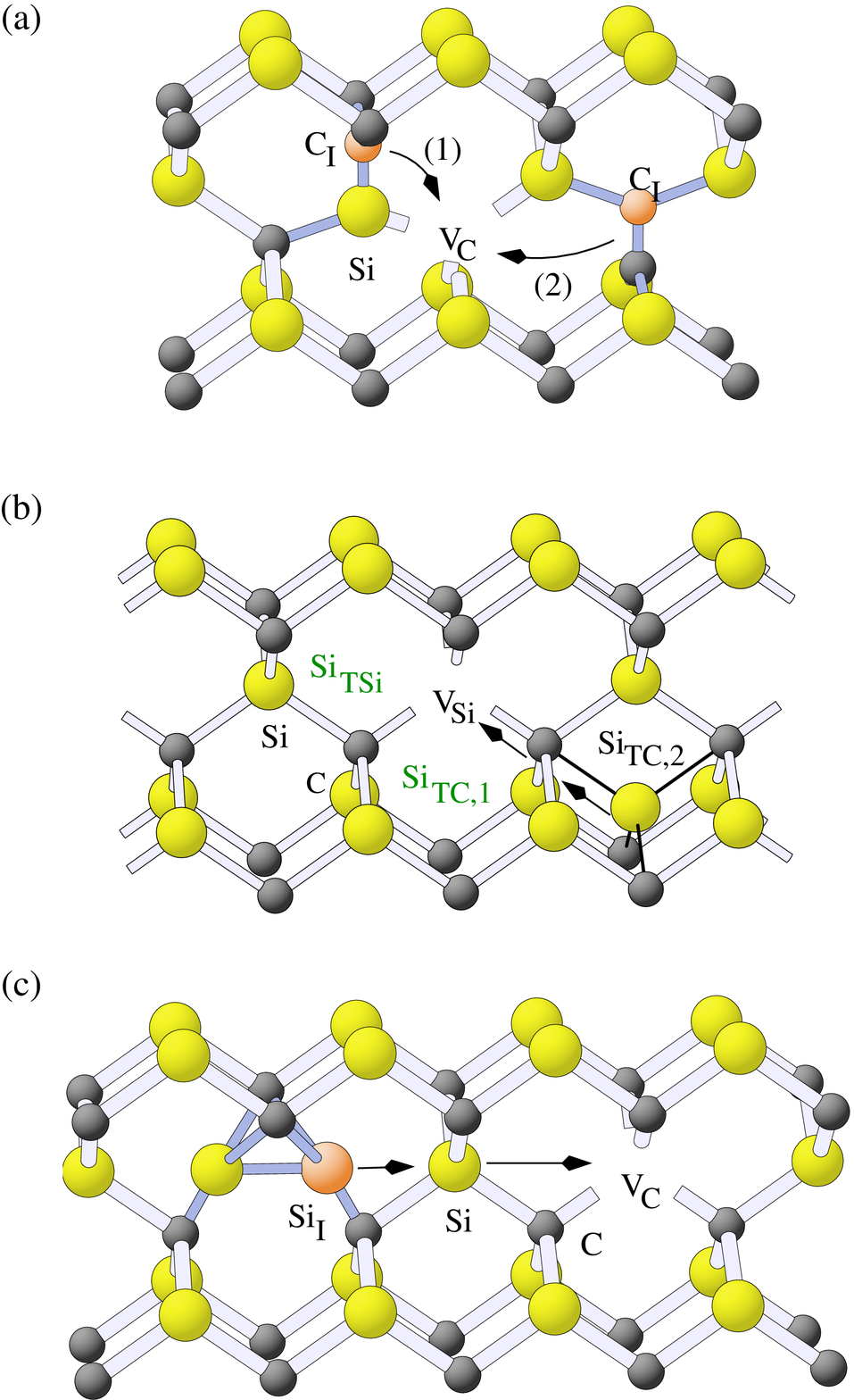}
\caption{\label{fig:i-v-pairs}Recombination of carbon and silicon Frenkel pairs: 
(a) carbon nearest-neighbor-pair
C$_{\mathrm{sp\/Si}\langle100\rangle}$-V$_{\mathrm{C}}$ 
and  second-neighbor pair
C$_{\mathrm{sp}\langle100\rangle}$-V$_{\mathrm{C}}$, (b) silicon Frenkel pairs
with Si$_{\mathrm{T\/C}}$-interstitial (the labels Si$_{\mathrm{T\/Si}}$ and
Si$_{\mathrm{T\/C,\,1}}$ indicate the unstable sites)
and (c)
fourth-neighbor-pair
Si$_{\mathrm{sp}\langle110}$-V$_{\mathrm{Si}}$. The lowest energetic path of the
interstitial atom is indicated by arrows.}
\end{figure}

The nearest-neighbor-pair
C$_{\mathrm{sp\/Si}\langle100\rangle}$-V$_{\mathrm{C}}$ is a vacancy-complex
with a carbon split-interstitial at the silicon site.  Two different
configurations are possible: a stable configuration with a threefold
coordinated carbon interstitial and a silicon dangling bond as depicted in
Fig~\ref{fig:i-v-pairs}a, and another configuration in which the carbon
interstitial possesses one carbon neighbor.  This latter pair turned out to be
unstable. The first configuration exists in the charge states 2$^{+}$ through
1$^{-}$ with the ionization levels at 0.4\,eV $(+|2+)$, 0.7\,eV $(0|+)$, and
1.8\,eV $(-|0)$. The defect states are formed by a silicon $p$-orbital 
of the interstitial and the dangling bond of the vacancy within the same
plane. Using the saddle point search method and a 64 atom cell with special
k-points we have obtained a recombination barrier for the neutral charge state
of 1.4\,eV (c.f.  Tab.~\ref{tab:i-v-pairs}).  At the transition state the
occupied defect levels are well localized. Towards the end of the
recombination they transform into delocalized valence band states.  The pair
may also recombine in the positive charge states. There we expect the
neutralization by an electron transfer at the end of the recombination.  In
this case a recombination barrier can be calculated and we obtain 1.2\,eV for
(C$_{\mathrm{sp\/Si}\langle100\rangle}$-V$_{\mathrm{C}}$)$^{2+}$ and
(C$_{\mathrm{sp\/Si}\langle100\rangle}$-V$_{\mathrm{C}}$)$^{1+}$.  For
(C$_{\mathrm{sp\/Si}\langle100\rangle}$-V$_{\mathrm{C}}$)$^{-}$, which is
relevant in n-type material, a second defect level is occupied that transforms
into a delocalized conduction band state during the recombination. Two
possibilities exist in this case: firstly, the defect is converted to neutral
by an electron transfer before the recombination or secondly, in the final
stage of the recombination the additional electron is donated into the
conduction band. The treatment of both cases is beyond the scope of the
present work.

In the second-neighbor-pair
C$_{\mathrm{sp}\langle100\rangle}$-V$_{\mathrm{C}}$ the interstitial is located
at the neighboring carbon site, as depicted in
Fig.~\ref{fig:i-v-pairs}a.  The pair exists in the charge states $2^{+}$,
1$^{+}$, 0 and 1$^{-}$, with the ionization levels
$\varepsilon(+|2+)$=0.8\,eV, $\varepsilon(0|+)$=1.1\,eV and
$\varepsilon(-|0)$=1.6\,eV.  The defect states mainly derive from the
interstitials levels. The recombination occurs by a second-neighbor hop of the
upper carbon atom via the unstable
C$_{\mathrm{sp\/Si}\langle100\rangle}$-configuration as indicated in
Fig.~\ref{fig:i-v-pairs}a. The recombination barriers for the pair are listed
in Tab.~\ref{tab:i-v-pairs}.  With values between 1.0\,eV and 0.4\,eV they
are considerably lower than the barriers for the recombination of
C$_{\mathrm{sp\/Si}\langle100\rangle}$-V$_{\mathrm{C}}$. The barrier is also
lower than the migration barrier of the split-interstitials.  For the negative
charge state a similar mechanism as for the negative nearest-neighbor pair has
to be considered.

In p-type material the vacancy and the approaching interstitial may be
positively charged. An additional long range Coulomb repulsion may hinder
the recombination.  The barrier for the formation of the Frenkel pair
is then given by the migration barrier of the interstitial plus the Coulomb
repulsion of the pair. On the other hand, once the defect levels begin to overlap new
states are formed. The charge state will be successively reduced by
electron transfer to the defect levels.  Indeed, for the Frenkel pairs
discussed above the total charge of the isolated defects
C$_{\mathrm{sp}\langle100\rangle}^{2+}$ and V$_{\mathrm{C}}^{2+}$ is reduced to
2$^{+}$ or 1$^{+}$. 

A considerable energy gain is associated with the formation of the
Frenkel pairs C$_{\mathrm{sp\/Si}\langle100\rangle}$-V$_{\mathrm{C}}$ and
C$_{\mathrm{sp}\langle100\rangle}$-V$_{\mathrm{C}}$ from the isolated defects
C$_{\mathrm{sp}\langle100\rangle}$ and V$_{\mathrm{C}}$.  The binding energy for
C$_{\mathrm{sp\/Si}\langle100\rangle}$-V$_{\mathrm{C}}$ varies between 2\,eV for
(C$_{\mathrm{sp\/Si}\langle100\rangle}$-V$_{\mathrm{C}}$)$^{2+}\rightarrow$C$_{\mathrm{sp\langle100}\rangle}^{0}$+V$_{\mathrm{C}}^{2+}$
and 4\,eV for
(C$_{\mathrm{sp\/Si}\langle100\rangle}$-V$_{\mathrm{C}}$)$^{0}\rightarrow$C$_{\mathrm{sp\langle100}\rangle}^{0}$+V$_{\mathrm{C}}^{0}$.
For the second-neighbor pair
C$_{\mathrm{sp}\langle100\rangle}$-V$_{\mathrm{C}}$ we obtain values between
0.8\,eV for (C$_{\mathrm{sp}\langle100\rangle}$-V$_{\mathrm{C}}$)$^{2+}\rightarrow$C$_{\mathrm{sp\langle100}\rangle}^{0}$+V$_{\mathrm{C}}^{2+}$
 and 1.0\,eV for the separation of the neutral pair. The interstitial is thus
attracted by the vacancy once the charge state has equilibrated.

Recently, the recombination of vacancies with carbon split-interstitials was
investigated in 4H-SiC using a DFT-based tight-binding scheme.\cite{rauls:01} 
Only neutral Frenkel pairs were considered. A similar recombination barrier of the neutral pair
C$_{\mathrm{sp\langle100\rangle}}$-V$_{\mathrm{C}}$ (0.5\,eV) was found as in our
calculation for 3C-SiC. Larger barriers were obtained for the direct
recombination of more remote pairs. 

\subsection{Silicon vacancies}
Frenkel pairs involving the silicon vacancy may be formed with either of the dominant
interstitials Si$_{\mathrm{T\/C}}$, Si$_{\mathrm{sp}\langle110\rangle}$ and
with the silicon-coordinated interstitial Si$_{\mathrm{T\/Si}}$.  The Frenkel pairs with
Si$_{\mathrm{T\/C}}$ and Si$_{\mathrm{T\/Si}}$ should account for the
recombination in p-type material, when the Si$_{\mathrm{T\/C}}$-site is
relevant for the migration. A recombination via the Frenkel pair with
Si$_{\mathrm{sp}\langle110\rangle}$ should be relevant in intrinsic
(compensated) and n-type material, where split-interstitials dominate.

First we have analyzed Frenkel pairs formed with a Si$_{\mathrm{T\/Si}}$- or
Si$_{\mathrm{T\/C}}$-interstitial neighboring the silicon vacancy. The Frenkel
pairs with a Si$_{\mathrm{T\/C}}$-interstitial or a
Si$_{\mathrm{T\/Si}}$-interstitial next to the vacancy turned out to be
unstable in all relevant charge states (the labels Si$_{\mathrm{T\/C,\,1}}$
and Si$_{\mathrm{T\/Si}}$ indicate the sites of the interstitial in
Fig.~\ref{fig:i-v-pairs}b). Upon reaching these sites, the interstitial
immediately recombines with the vacancy.  The closest stable Frenkel pair we
found involves an interstitial at a Si$_{\mathrm{T\/C}}$-site with only one
common carbon neighbor to the vacancy (c.f. the Si$_{\mathrm{T\/C,\,2}}$-site
in Fig.~\ref{fig:i-v-pairs}b). This stable Frenkel pair has deep levels within
the band gap. It is only realized in the positive and neutral charge state
with ionization levels at $\varepsilon({+|2+})=0.5$\,eV and
$\varepsilon({0|+})=1.2$\,eV.  The recombination of this pair could be
achieved via a hop of the interstitial passing close by the adjacent
Si$_{\mathrm{T\/Si}}$-site.  Our calculations indicate that this is not a
likely path. Instead an anti-structure pair forms by a hop of the carbon
neighbor into the silicon vacancy. In a concerted motion the interstitial
moves into the site left by the carbon neighbor. The reaction barrier amounts
to 3.2\,eV for the positive and the neutral Frenkel pair, which is comparable
with the migration barrier of Si$_{\mathrm{T\/C}}$.

Frenkel pairs with the split-interstitial Si$_{\mathrm{sp}\langle110\rangle}$
represent other possible configurations. It turns out that
Si$_{\mathrm{sp}\,\langle110\rangle}$-V$_{\mathrm{Si}}$ is unstable when the
vacancy and the interstitial are second neighbors, irrespective of the
orientation of the pair.  A recombination barrier thus may be encountered for
more distant pairs, e.g. with the interstitial at the third or fourth neighbor
shell. The recombination of such pairs is based on the migration of the split
interstitial to an unstable second neighbor position. Here we discuss the
fourth-neighbor Frenkel pair in more detail. Our findings for this pair
suggest a similar mechanism for the third neighbor pair. The
fourth-neighbor complex is stable except in highly doped p-type material. In
the doubly positive charge state the interstitial and the vacancy immediately
recombine.  The complex becomes positively charged at the ionization levels
(2$^{+}|1^{+}$)=0.48\,eV.  For a Fermi level above mid-gap the complex is
neutral or negative (ionization levels: (1$^{+}|$0)=1.3\,eV and
(0$|1^{-}$)=1.8\,eV). It is doubly negative in strongly doped n-type material
only (ionization level(2$^{-}|1^{-}$)=2.3\,eV). The saddle point search finds
the following recombination mechanism (c.f. Fig.~\ref{fig:i-v-pairs}c): first
the silicon interstitial jumps towards the silicon-neighbor it has in common
with the vacancy, then this neighbor is kicked out of its site and recombines
with the vacancy. This process is accompanied with a barrier of 0.2\,eV. As
the split-interstitial close to the vacancy is unstable, the barrier for this
process is considerably lower than the barrier for the interstitial migration.
As discussed for the annealing of carbon vacancies, we expect a higher barrier
for the recombination in the negative charge state of the complex.

Thus the analysis of the silicon Frenkel pairs shows that the pairs with a short distance
between the interstitial and the vacancy are unstable, except for the second
neighbor pair pair V$_{\mathrm{Si}}$-Si$_{\mathrm{T\/C,\,2}}$. For the stable pairs we
have found recombination barriers that are somewhat lower than the migration barriers
of the interstitial. These findings suggest a recombination based
on a migration towards the closest stable sites. They also indicate that the
energy barriers of this process should not exceed the migration barrier of
the corresponding interstitial.
\section{Clustering of carbon interstitials}
\label{sec:cluster}
\begin{figure*}
\includegraphics[width=\linewidth]{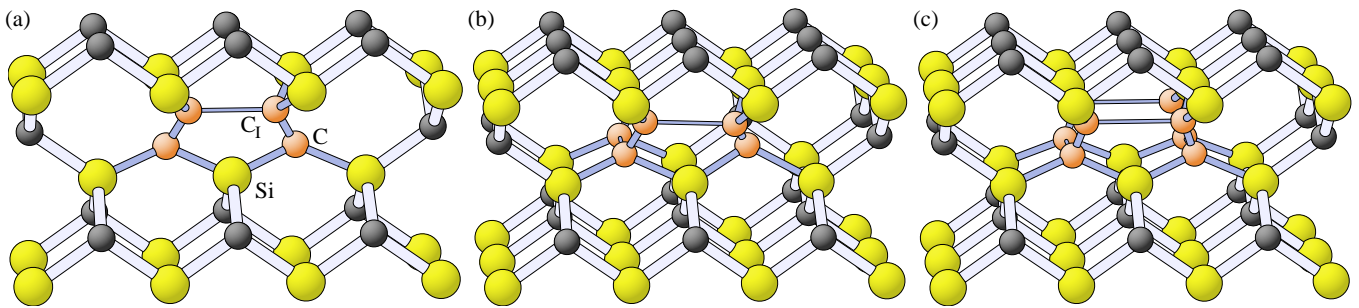}
\caption{\label{fig:3C-cluster} Carbon insterstitial
  clusters in 3C-SiC: (a) Di-interstitial (C$_{\text{sp}}$)$_{2}$
  (b) Tri-interstitial (C$_{\text{sp}}$)$_{3}$ (c) Tetra-interstitial
  (C$_{\text{sp}}$)$_{4}$}
\end{figure*}

\begin{figure}
  \includegraphics[width=0.7\linewidth]{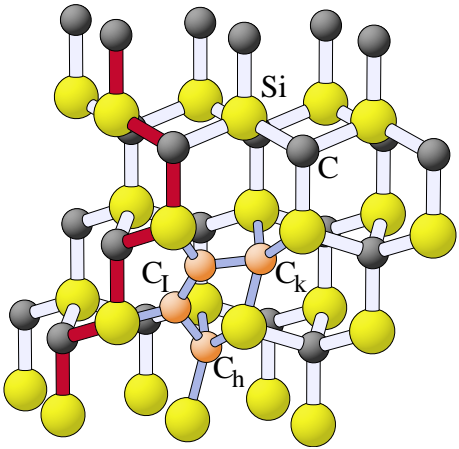}
  \caption{Di-interstitial (C$_{\text{sp}}$)$_{2, \text{kh}}$ in
    4H-SiC located on a cubic and a hexagonal site. The subscripts $k$ and $h$
    refer to cubic and hexagonal sites. $C_{\mathrm{I}}$ refers to the additional
    carbon atoms. The stacking sequence
    of the crystal is indicated by the dark bonds.}
  \label{fig:4H-cluster}
\end{figure}

\begin{table}
  \caption{  \label{tab:cluster-bind} Dissociation energy of
    neutral interstitial and vacancy clusters. For the interstitial
    clusters the given value is the energy needed to remove a single
    carbon atom.}
  \begin{ruledtabular}
    \begin{tabular}{ccc}
      Polytype & Cluster & Dissociation energy (eV) \\ 
      \hline
      3C & (C$_{\text{sp}}$)$_{2}$ & 2.8 \\
      3C & (C$_{\text{sp}}$)$_{3}$ & 3.0 \\
      3C & (C$_{\text{sp}}$)$_{4}$ & 5.7 \\
      4H & (C$_{\text{sp}}$)$_{2, \text{kh}}$ & 5.8 \\
      4H & (C$_{\text{sp}}$)$_{2, \text{hh}}$ & 5.0 \\
      4H & (C$_{\text{sp}}$)$_{2, \text{kk}}$ & 4.6 \\
      \hline
      3C & V$_{\text{C}}$-V$_{\text{Si}}$ & 4.5 \\
      3C & V$_{\text{C}}$-V$_{\text{C}}$ & 1.7 \\
      3C & V$_{\text{Si}}$-V$_{\text{Si}}$ & 0.1 \\
    \end{tabular}
  \end{ruledtabular}
\end{table}

\begin{table}
  \caption{  \label{tab:cluster-charge}
  Ionization levels of the investigated interstitial clusters. The
  tri-interstitial (C$_{\text{sp}}$)$_{3}$ exhibits a negative-$U$
  effect for the positive charge states and the tetra-interstitial
  (C$_{\text{sp}}$)$_{4}$ is electrically inactive.}
  \begin{ruledtabular}
    \begin{tabular}{ccccccc}
      &&$\varepsilon(1^{+}|2^{+})$&$\varepsilon(0|1^{+})$&$\varepsilon(1^{-}|0)$&$\varepsilon(2^{-}|1^{-})$\\
      \hline
      3C & (C$_{\text{sp}}$)$_{2}$ & 0.8  & 0.99 & - & - & \\
      3C & (C$_{\text{sp}}$)$_{3}$ & 0.85 & 0.62 & 1.49  & 1.8 & \\
      3C & (C$_{\text{sp}}$)$_{4}$ & - & - & - & - & \\
      \hline
      4H & (C$_{\text{sp}}$)$_{2, \text{kh}}$ & -    & 0.10 & 2.78 & 3.13 & \\
      4H & (C$_{\text{sp}}$)$_{2, \text{hh}}$ & 0.41 & 0.50 & 2.82 & 2.68 & \\
      4H & (C$_{\text{sp}}$)$_{2, \text{kk}}$ & 0.32 & 0.38 & 2.66 & 2.6 & \\
    \end{tabular}
  \end{ruledtabular}
\end{table}

A formation of composite defects, e.g. D$_{\text{I}}$ or
D$_{\text{II}}$ centers, during annealing requires a reservoir of
intrinsic defects. However, at elevated temperatures vacancies and
interstitials anneal out. (see discussion in Sec.~\ref{sec:discuss}).
Yet, the point defects can be kinetically supplied by defect
precipitations. The clusters of vacancies and interstitials can then
serve as sources of mobile defects. We expect that the effect of the
interstitial clusters on the defect kinetics is much more pronounced
than that of the vacancy clusters. Although both cluster types possess
sizeable dissociation energies (cf.  Tab.~\ref{tab:cluster-bind} and
Ref.~\onlinecite{torpo:02} for the divacancy), the migration barriers
of the carbon and the silicon vacancies are much higher than those of
the interstitials (cf.  Sec.~\ref{sec:v.i.mig}) so that it is much
more likely that the interstitials combine into precipitates. In
addition, we expect (cf.  Sec.~\ref{sec:discuss}) that the
vacancy-interstitial recombination sets in before a significant
vacancy clustering occurs.

We focus in this section on small clusters formed by upto four carbon
interstitials. Due to the high mobility of the carbon interstitials,
these clusters may act as a sink for interstitials at lower
temperatures and can re-emit them at higher temperatures. Similar
aggregates have already been investigated in diamond.\cite{goss:01}
The microscopic structure of the considered complexes is displayed in
Fig.~\ref{fig:3C-cluster} for 3C-SiC, an example of the
di-interstitial in 4H-SiC is shown in Fig.~\ref{fig:4H-cluster}. A
detailed analysis of the electronic structure and the vibrational
spectra is given elsewhere.\cite{mattausch:03a} Here we summarize the
facts that are important for the annealing mechanisms. The focus is on
the dissociation energies of these defects, i.e.\ the energy needed to
remove a single carbon atom from these clusters.  For a mid-gap
Fermi-level all clusters are neutral. The results for the dissociation
energy in the neutral charge state are listed in
Tab.~\ref{tab:cluster-bind}.

The carbon split-interstitial possesses two dangling $p$-orbitals
oriented in $\langle110\rangle$-direction, resulting from the
sp$^{2}$-hybridization of the carbon atoms. The two adjacent
split-interstitials form a covalent bond due to the overlap of these
orbitals. This (cf. Fig.~\ref{fig:3C-cluster}a) results in a more
favorable sp$^{3}$-hybridization of the two interstitial carbon atoms
and gives rise to an energy gain of 2.8\,eV for the carbon
di-interstitial (C$_{\mathrm{sp}}$)$_{2}$.  In the same way further
carbon atoms can be absorbed, resulting in larger carbon aggregates.
The tri-interstitial (Fig.~\ref{fig:3C-cluster}b) possesses a similar
dissoziation energy (3.0\,eV) as the di-interstitial. In the
tetra-interstitial a ring of split-interstitials is formed
(Fig.~\ref{fig:3C-cluster}c) which is especially favorable. All carbon
atoms are sp$^{3}$-hybrids, hence all bonds are saturated.  This leads
to an energy gain of 5.7\,eV for the added carbon interstitial.

The charge states of the investigated clusters are listed in
Tab.~\ref{tab:cluster-charge}.  The di-interstitial possesses all
charge states from $2+$ to $0$.
The tri-interstitial is mainly neutral, only at the band edges the
charge states $2+$ and $2-$ can exist. This cluster exhibits a
negative-$U$ effect, since it is energetically favorable to fully
occupy the connecting bonds between all three adjacent
split-interstitials (cf.  Fig.~\ref{fig:3C-cluster}b).  The
tetra-interstitial is electrically inactive, since all carbon
$p$-orbitals are saturated.

In 4H-SiC, a stronger relaxation and rehybidization of the carbon
atoms is observed than in 3C-SiC. Therefore already the
di-interstitial shows a very high stability.  Three different
variations of the di-interstitial are possible. The
split-interstitials can either lie completely in a hexagonal or a
cubic plane, or they occupy two adjacent cubic and hexagonal sites
(cf. Fig.~\ref{fig:4H-cluster}).  With a binding energy of 5.8\,eV the
latter is the most stable configuration. This higher binding energy
compared to 3C-SiC results from the sp$^{3}$-hybridization of the
border atoms, which is preferred due to the different geometry of the
hexagonal planes in 4H-SiC.  Also the hexagonal and cubic
di-interstitials are very stable with binding energies of 5.0\,eV and
4.6\,eV, respectively. These di-interstitials show a strong distortion and have
a nearly linear structure within their cubic or hexagonal plane.

The micro-clusters presented in this section are examples of clustering processes.
The formation of cluster networks via bonds between the p-orbitals of adjacent
$(C_{\mathrm{sp}})_{2}$ and $(C_{\mathrm{sp}})_{3}$ clusters is also possible.
Additionally, carbon antisites can trap carbon interstitials with sizable
binding energies.\cite{mattausch:01} In the discussion of the annealing
mechanisms below we outline the competition of the cluster formation with the
vacancy-interstitial recombination and their role as a soure for carbon
interstitials at the high temperature annealing stages.
\section{Hierarchy of annealing mechanisms}
\label{sec:discuss}
The annealing of vacancies and interstitials is governed by several competing
mechanisms.  The principal mechanisms are (i) the recombination of vacancies
and interstitials, (ii) the out-diffusion to the surface, (iii) the diffusion
to sinks, where they are annihilated or otherwise form stable complexes, and
(iv) the transformation into other more stable
configurations that is not diffusion-related.  Sinks for interstitials or
vacancies could be immobile defects, such as interstitial clusters and stable
complexes with impurities, or extended defects, such as dislocations.  In the
latter cases the interstitial or vacancy are not
disappearing as such, but are bound in a more stable form. In the experiment,
nevertheless, the observed signatures \emph{anneal out}, while other centers
related to the annealing may appear in the spectra.

The experimental investigation of the annealing kinetics requires a sufficient
temperature $T$ to activate one or more of the above annealing mechanisms. The
diffusion-limited mechanisms (ii)-(iii) and the transformation-related
mechanisms are typically described by first order kinetics with an activation
energy given by the migration or transformation barrier (e.g.
Ref.~\onlinecite{bourgoin:83}). The kinetics of the vacancy-interstitial
recombination depends on the average separation of the vacancy and
interstitial. For bound Frenkel pairs the activation energy is given by the
recombination barrier. For uncorrelated vacancy-interstitial pairs (e.g.  when
their separation is much larger than the capture radius) the interstitial has
to migrate to the vacancy (the vacancies are rather immobile) to form a
Frenkel pair. The Frenkel pair then simply recombines by a hop of the
interstitial into the vacant lattice site.  In this case, as the recombination
barriers of the Frenkel pair are found to be relatively small, the
recombination is essentially limited by the diffusion of the interstitial. The
recombination then is described by the activation energy of the interstitial
migration.  The prefactors of the diffusion-limited mechanisms and the
vacancy-interstitial recombination, however, depend on parameters arising from
the sample prepartion, e.g. the average distribution (and concentration) of
the defects (and sinks), as well as the mutual interaction between the
involved partners.  Hence, the predictive translation of activation energies
into annealing temperatures hinges on parameters that are specific to the
particular samples. Only by detailed simulations it is possible
to assess the initial state of the samples (e.g. after irradiation) and the
annealing kinetics quantitatively. Our analysis forms the basis for such
simulations, that are well beyond the scope of the present paper.
Nevertheless, a qualitative assessment of the kinetics via the calculated
activation energies should be valid, since the activation energies differ
substantially in value and enter the rate constants exponentially.

In the following we deduce a hierarchy of annealing mechanisms based on a
comparison of the energy barriers involved in the principal annealing
mechanisms. Central to the discussion are four important observations.  Our
results for the migration of intrinsic point defects show that the
interstitials are by far more mobile than the vacancies, also the recombination
barriers for Frenkel pairs are lower than the migration barriers of
the vacancies. The silicon vacancy is a metastable defect in p-type and
compensated material. The migration of defects strongly depends on the charge state,
in particular the migration of silicon interstitials and vacancies.

\subsection{Silicon vacancies and interstitials}
Qualitatively, the annealing of silicon vacancies is subject to three
different Fermi-level dependent mechanisms: (i) the transformation of the
silicon vacancy into a carbon vacancy-antisite complex, (ii) the Frenkel pair
recombination with silicon split-interstitials and (iii) the migration of silicon
vacancies to sinks. These mechanisms are summarized in Fig.~\ref{fig:scn.anneal}.

In p-type material, we expect a dominance of the annealing of the vacancy
signature by its transformation into the vacancy-antisite complex.  The
recombination of the stable Frenkel pairs, i.e. pairs with the
Si$_{\mathrm{T\/C}}$-interstitial located in the second neighbor shell or
beyond, is suppressed by the larger migration/recombination barrier
(3.2\,eV-3.5\,eV) as compared to the barrier of the transformation
(1.9\,eV-2.4\,eV). Also the migration of the silicon vacancy is kinetically
suppressed in p-type material due to the larger migration barrier
(3.6\,eV-3.2\,eV). Once the silicon vacancy is transformed into the
vacancy-antisite complex, it stays in this configuration. The reverse
transformation requires fairly high temperatures due to the large barrier of
6.5\,eV. As discussed in Sec.~\ref{sec:meta}, the dissociation of the
vacancy-antisite complex becomes possible at the same time.

In compensated material (with the Fermi level at or above mid-gap), the
silicon split-interstitial is the most relevant interstitial
configuration. Its migration barrier is lower than the
transformation barrier of the silicon vacancy. Hence the vacancy-interstitial
recombination prevails in the annealing hierarchy.  For the recombination of
bound Frenkel pairs we found a low barrier of 0.2\,eV. The
annihilation of vacancies by more distant interstitials is limited by the
migration of the interstitial towards the vacancy with a migration barrier of
1.4\,eV. It constitutes a separate annealing stage. Note that the small
binding energy of the Frenkel pair indicates a weak attractive
interaction between the interstitial and the vacancy. Hence a dissociation of
distant pairs has a relatively high probability, which enables further annealing
stages.  The metastability-induced annealing is the next annealing
stage in the hierarchy. As a final stage, the vacancy-antisite
complex, that evolves from the silicon vacancy, anneals by dissociation or by
its migration to sinks. Again the migration of silicon vacancies does not
directly contribute as the migration barrier of 3.2\,eV to 3.4\,eV (V$_{\mathrm{Si}}^{-}$
and V$_{\mathrm{Si}}^{0}$ respectively) is larger than the transformation barrier.

In n-type material the silicon vacancy becomes a stable defect. As in
intrinsic (compensated) material the annealing by vacancy-interstitial
recombination has a lower activation barriers than the vacancy migration. The
migration of silicon vacancies to sinks is most likely the last stage in the
hierarchy.  It has a slightly higher barrier than the transformation
V$_{\mathrm{Si}}\rightarrow$V$_{\mathrm{C}}$-C$_{\mathrm{Si}}$ in
compensated material.

By electron spin resonance experiments Itoh \emph{et~al.}~\cite{itoh:97a}
observed the annealing of the T1-center in p-type and n-type 3C-SiC irradiated
by electrons and protons.  They found three annealing stages at
150$^{\circ}$C, 350$^{\circ}$C and 750$^{\circ}$C. The annealing behavior was
found to be insensitive to the doping (aluminum and nitrogen) and the
irradiating particles (electrons and protons), except for the p-type
electron-irradiated sample, where the final annealing occurred at
350$^{\circ}$C in coincidence with the annealing of three other
defect-centers. Itoh \emph{et~al}. argued that a possible loss of the
paramagnetic state might be correlated with the annealing of two unidentified
centers T6 and T7 that were only observed under these
conditions.\cite{itoh:97a}. The independence of the annealing behavior of the
particular dopants indicates that the observed annealing process itself
involves only intrinsic defects.

The T1-center was identified by the authors as a negative silicon vacancy
V$_{\mathrm{Si}}^{-}$. Later this identification was verified
theoretically\cite{wimbauer:97,petrenko:01,bockstedte:02,bockstedte:03b} by
calculations of the hyperfine tensors.  According to earlier theoretical
results\cite{wimbauer:97,zywietz:99,torpo:01} and our calculations
(c.f.~Tab.~\ref{tab:ion}) the silicon vacancy is in a negative charge state
over a wide range of the Fermi level starting below mid-gap
($\mu_{\mathrm{F}}>$0.6\,eV) and ending somewhat below the conduction band
edge ($\mu_{\mathrm{F}}<$1.76\,eV). This suggest that the p-type and n-type
samples were compensated by the irradiation, most likely the Fermi level was
trapped by irradiation-induced deep defects around mid gap. Under these
conditions the scenario for compensated material we described above should
explain the annealing of the EPR-signature of V$_{\mathrm{Si}}^{-}$ by three
annealing stages, namely: (i) the initial recombination of Frenkel pairs with
a small separation between the vacancy and interstitial, (ii) the
recombination of vacancies and interstials with a larger separtion that is limited by the diffusion
of the interstitial, and (iii) the transformation of V$_{\mathrm{Si}}^{-}$
into a carbon vacancy-antisite complex.

From the annealing experiments Itoh~\emph{et~al}.\cite{itoh:89} deduced an
activation energy of 2.2\,eV for the last annealing stage. According to our
calculations the transformation of V$_{\mathrm{Si}}$ into
V$_{\mathrm{C}}$-C$_{\mathrm{Si}}$, that constitutes the third annealing
stage, is associated with an activation energy of 2.5\,eV
(c.f.~Tab.~\ref{tab:v.mig}) in good agreement with the experimental result.
Once this transformation is activated silicon vacancies are largely
transformed into carbon vacancy-antisite complexes and the EPR-signal of the
T1-center vanishes. During the short observation times (5\,min) of the
isochronal annealing the reappearance of V$_{\mathrm{Si}}^{-}$ is kinetically
hindered.  Even though a signature of the carbon vacancy-antisite complex may
develop in the spectra, Itoh \emph{et~al}.  reportedly have not observed the
appearance of a new defect center that could evolve from the silicon vacancy.
Note, that for a Fermi level below 1.24\,eV the vacancy-antisite complex is
not paramagnetic. Only in a small window, for a Fermi level between 1.24\,eV
and 1.79\,eV the paramagnetic positive charge state should prevail.  Recently,
Lingner~\emph{et~al}. identified the
V$_{\mathrm{C}}$-C$_{\mathrm{Si}}$-complex by spin resonance experiments in
neutron-irradiated n-type 6H-SiC.\cite{lingner:01} The hyperfine structure of
an excited high spin state ($S=1$) of this complex was detected using the
magnetic dichroism of the adsorption (MCDA) and MCDA-EPR.  The theoretical
analysis showed that the excited state is related to an intra-defect
excitation of (V$_{\mathrm{C}}$-C$_{\mathrm{Si}}$)$^{2+}$, which is not
paramagnetic in its ground state and therefore could not be detected by
standard EPR. This is consistent with the observation of Itoh \emph{et~al}.
The complex appeared only in samples annealed above the temperature at which
the T1-center in 3C-SiC finally vanishes.  EPR-centers with a similar
$g$-tensor and fine structure constant $D$ were observed earlier in n-type
6H-SiC\cite{vainer:81} (P6- and P7-centers) and in electron irradiated
nominally undoped 3C-SiC (L3-center).\cite{son:97} The similarity suggests
that the carbon vacancy-antisite complex as a common model for all these
centers.  Indeed, the annealing study of Son \emph{et~al}. shows that the
center appears only after an annealing at 700$^{\circ}$C-750$^{\circ}$C.
These observation support our interpretation of the annealing stages for the
silicon vacancy reported by Itoh \emph{et~al}.. An diffusion-based annealing
mechanism, that was proposed earlier,\cite{itoh:97a,son:97} is apparently not
involved in these experiments.

Using PAS Kawasuso \emph{et~al}.~\cite{kawasuso:96,kawasuso:98} have observed
the annealing of vacancy-related defects in electron-irradiated n-type 3C-SiC
and 6H-SiC. Two lifetime components were extracted from the positron signal in
6H-SiC. One component was attributed to defects related to silicon
vacancies.\cite{brauer:96a,staab:01} The other was shown to originate from
carbon vacancy-related defects. In 3C-SiC only the component corresponding to
silicon vacancy-related defects was detected. Kawasuso \emph{et~al}. found
that this lifetime component annealed in several stages.  For annealing
temperatures below 500$^{\circ}$C two annealing stages were observed in 3C-
and 6H-SiC that followed the findings for the T1-center in 3C-SiC. In 3C-SiC
vacancy-related defects were not detected above the annealing temperature of
the T1-center.  This finding is also consistent with our interpretation of the
annealing stages, since a positive carbon vacancy-antisite complex would repel
the positron and therefore should not contribute to the positron lifetime.

In 6H-SiC, however, the signal persisted the annealing at 750$^{\circ}$C and
finally annealed at 1450$^{\circ}$C.  Yet, an annealing stage with a less
pronounced drop of the vacancy concentration is present at 750$^{\circ}$C.  In
a more recent PAS-study~\cite{kawasuso:01} using electron irradiated n-type
6H-SiC Kawasuso~\emph{et~al}. describes the annealing of the silicon vacancy or
related defect complexes with the a major stages at
500$^{\circ}$C-700$^{\circ}$C and 1000$^{\circ}$C-1200$^{\circ}$C. The
vacancy-signature reached the bulk-level at 1200$^{\circ}$C. In comparison to
this more recent study, the samples of the earlier study had a larger nitrogen
concentration (carrier density of $5.5\times10^{17}$\,cm$^{-3}$ at room
temperature vs.  $5\times10^{15}$\,cm$^{-3}$) and were irradiated with more
energetic electrons at a lower dose (1$\times10^{17}\,e^{-}$/cm$^{2}$ at
3\,MeV compared 3$\times10^{17}\,e^{-}$/cm$^{2}$ at 2\,MeV and). The
differences in the annealing behavior is most likely related to these
variation in the nitrogen concentrations and the irradiation conditions. 

In the following we comment on presence of the annealing stages above
1000$^{\circ}$C in 6H-SiC. In their analysis of the earlier experiment Kawasuso
\emph{et~al}. explained the final annealing stage at 1450$^{\circ}$C in terms
of the dissociation of vacancy-nitrogen complexes and the subsequent annealing
of the vacancy by out-diffusion or diffusion to sinks. The V$_{\mathrm{Si}}$-N
complexes were expected to form during the annealing. Such an interpretation
is plausible as similar positron lifetimes are predicted for the
V$_{\mathrm{Si}}$-N complexes and the silicon vacancy.\cite{brauer:96a} Very
recent theoretical investigations show that these complexes are thermally
stable in agreement with EPR-experiments.\cite{gerstmann:03,vainer:81a} Yet,
also the transformation into the carbon vacancy-antisite complex may play a
relevant role in the PAS experiments, provided that the sample remains
compensated (Fermi level around 1.7\,eV) and that the positron annihilation at
the carbon vacancy-antisite complex resembles that for the silicon vacancy.
Precise calculations based on \emph{ab initio} methods for this center are
lacking sofar.  However, calculations\cite{kawasuso:privcom} using an
approximate method based on the superposition of atomic densities (c.f.
Ref.~\onlinecite{puska:94}) using the atomic coordinates obtained for
V$_{\mathrm{C}}$-C$_{\mathrm{Si}}$ obtained in our work indicate for 4H-SiC,
that the positron lifetime associated with this center lie in the range
similar to the value of the silicon vacancy (at the cubic site between 162\,ps
to 175\,ps and at the hexagonal site 187\,ps to 193\,ps depending on the
charge state). Thus the positron lifetime does not significantly change when the
negative silicon vacancy transforms into the neutral hexagonal
vacancy-antisite complex. The migration of the carbon vacancy-antisite complex
is associated with larger barriers than the transformation and hence is
activated at higher temperatures.  This scenario is a plausible explanation of
the observed annealing of the V$_{\mathrm{Si}}$-related signature in the PAS
experiments, since the complex at the hexagonal site may become neutral during
the annealing of compensating centers in n-type 4H-SiC. Obviously, Fermi level
effects are relevant. The annealing characteristics should therefore depend on
the polytype (due to the different band gap width).  Based on our present
understanding, we do not exclude the formation of vacancy-nitrogen complexes
as an explanation of the final annealing stage.  The interpretation of the
last annealing stage in terms of vacancy-nitrogen complexes, however,
requires that the dopant is present in similar or higher concentrations than
the vacancies at the onset of the annealing stage.

\subsection{Carbon vacancies and interstitals}
For the carbon vacancy the annealing hierarchy starts with the
vacancy-interstitial recombination at the first annealing stage and ends with the
diffusion of vacancies to sinks or its out-diffusion (c.f.  Fig.~\ref{fig:scn.anneal}).  The
recombination of Frenkel pairs containing the carbon
split-interstitial as the nearest or second neighbor of the vacancy proceeds with
lower recombination barriers than the vacancy diffusion.  Similarly the
migration barrier for the carbon split-interstitial is lower than for the
carbon vacancy. For the vacancy-interstitial recombination, the migration of
the interstitial towards the vacancy is the bottle neck, since recombination
of the nearest or second neighbor Frenkel pairs proceeds with a similar or
lower barrier. The annealing of the vacancy based on its diffusion therefore
should be activated at much higher temperatures than the vacancy-interstitial
recombination.

In irradiated material, where interstitials and vacancies are present in
similar concentrations, we expect the vacancy-interstitial recombination to
represent the first annealing stage. Our finding of a relevant binding energy
for all charge states of the nearest and second neighbor Frenkel pair suggests
an attractive interaction that traps the interstitial in the vicinity of the
vacancy. Hence, we expect that a recombination of these Frenkel pairs has a
higher probability than their dissociation.  However, the diffusion-limited
recombination of the isolated defects may be hindered by a Coulomb repulsion
of the charged defects.  This occurs only in p-type material for a Fermi level
below 0.8\,eV (cf. Tab.~\ref{tab:ion}).  Above this value the carbon
split-interstitial is neutral.  Even though the split-interstitial becomes
negative in n-type 3C-SiC, the vacancy should remain neutral. For n-type
4H-SiC, on the other hand, our calculations show that also the carbon vacancy
becomes negative as a consequence of the larger band gap.  The additional
Coulomb barrier in p-type 3C-SiC and p-type or n-type 4H-SiC may leave a
considerable fraction of vacancies unannealed in this first annealing stage.
At the same time the carbon split-interstitials could anneal by diffusion to
other defects where thermally stable defects complexes or defect clusters
could be formed.  The remaining vacancies anneal in a second annealing stage
based on the vacancy diffusion.

So far the isolated carbon vacancy was not detected by EPR-experiments nor by
PAS experiments in irradiated 3C-SiC.  A spin
resonance center T5 was observed together with the T1-center
(V$_{\mathrm{Si}}^{-}$) by Itoh \emph{et al}. that annealed at 150$^{\circ}$C.
Its hyperfine (HF) signature originates from four silicon atoms and it
was therefore  identified as a defect on the carbon sublattice. However, the
original assignment to the positive carbon vacancy had been revised to a
complex of two hydrogen atoms with a carbon vacancy,\cite{aradi:01,gali:00}
given the $D_{2}$-symmetry of the center that is incompatible with the
theoretical findings\cite{zywietz:99,torpo:01} for the carbon vacancy.  Based
on the calculations of HF-parameters for these models and the carbon
split-interstitial it was recently shown that only the split-interstitial may
explain the experimental findings for the
T5-center.\cite{petrenko:02,bockstedte:03a,bockstedte:02,gali:03} Such an
identification of the T5-center has to be verified experimentally, as the
carbon HF-tensors of the $\langle100\rangle$-oriented carbon dumbbell have
sofar not been observed. Yet, this model would explain the low annealing
temperature of the T5-center. According to our analysis the positive carbon
split-interstitial is indeed a highly mobile defect, with a migration barrier
of only 0.9\,eV.  A comparison of the ionization levels of this defect and
V$_{\mathrm{Si}}$ (cf.  Tab.~\ref{tab:ion}) shows that
C$_{\mathrm{sp\langle100\rangle}}^{+}$ and V$_{\mathrm{Si}}^{-}$ simultaneous
exist in a paramagnetic state for a Fermi-level around 0.7\,eV in the p-type
samples which is in agreement with the simultaneous observation of the T1- and
T5-center.\cite{itoh:97a} At these doping conditions the carbon vacancy prevails in
the non-paramagnetic doubly positive charge state. This explains why the carbon
vacancy was not detected in EPR experiments or by PAS (the positron
recombination at the positive vacancy is suppressed by the Coulomb repulsion).
Therefore the mechanism (vacancy-interstitial recombination or
other diffusion-limited mechanisms) behind the annealing of the
T5-center cannot be deduced from the present experiments. 

A competing process to the vacancy-interstitial recombination is the formation
of carbon interstitial clusters as discussed in Sec.~\ref{sec:cluster}.  Due
to their high dissociation energies such clusters would emit interstitials
only at elevated annealing temperatures. These interstitials then take part in
further defect reactions. For the di-interstitial and the tri-interstitial in
3C-SiC a dissociation energy of 3\,eV plus an additional barrier, which may be
approximated by the migration barrier of the interstitial (0.5\,eV for
C$_{\mathrm{sp}\langle100\rangle}^{0}$), is needed to remove a carbon atom.
This activation barrier is only slightly lower than the migration barrier of
the carbon vacancy (4.1\,eV for V$_{\mathrm{C}}^{+}$ and 3.5\,eV for
V$_{\mathrm{C}}^{0}$). With a dissociation barrier of 5.7\,eV the
tetra-interstitial is much more stable and should emit interstitials at higher
temperatures than the di- and tri-interstitial.  A comparison of the
dissociation energies with the activation energies of the other processes
suggests that new carbon interstitials should be supplied at temperatures
above 1000$^{\circ{}}$C.  Indeed, the concentration of e.g.\ the
photo-luminescence centers D$_{\text{I}}$\cite{patrick:72a} and
D$_{\text{II}}$,\cite{patrick:73} which are presumably related to
antisites\cite{eberlein:02,gali:03a} and carbon
aggregates,\cite{patrick:73,mattausch:02} rises significantly at annealing
temperatures above 1000$^{\circ{}}$C.\cite{freitas:87,egilsson:99} For both
formation processes the availability of free carbon atoms is an important
pre-requisite.\cite{eberlein:02,mattausch:02} Defects similar to the carbon
clusters described here are known from other semiconductors.  Besides the
carbon clusters in diamond\cite{goss:01} one can also mention here the well known
\{311\}-interstitial defects in silicon. These defects emit silicon
interstitials and thereby drive the transient enhanced diffusion of dopants
during the annealing.\cite{eaglesham:94}

In contrast to 3C-SiC, the signatures of the carbon vacancy and interstial
were reported in 4H-SiC.
In irradiated p-type 4H-SiC, recently two EPR-centers (EI1 with S=1/2 and EI3
with S=1) were identified by Son \emph{et al}. (Ref.~\onlinecite{son:01b})
that possess a similar $g$-tensor as the T5-center in 3C-SiC. The analysis of
the HF-signature showed that the defects are located on the carbon sublattice.
Theoretical calculations indicate that the observed silicon HF-tensors are
consistent with the theoretical values for the positive and neutral carbon
split-interstitial, respectively.\cite{gali:03,bockstedte:03,petrenko:02} Similar to the
T5-center the EI1 and EI3 center anneal around 200$^{\circ}$C.  
In p-type 4H-SiC irradiated at 400$^{\circ}$C Son
\emph{et~al}.~\cite{son:01,son:01b} observed by EPR-experiments the
final disappearance of the EI5-center at 500$^{\circ}$C. The center was
tentatively assigned to a positive carbon vacancy. This assignment was verified
by calculations of HF-tensors,\cite{petrenko:01,bockstedte:02,bockstedte:03}
and it was shown that the center is located at the cubic site.  The low annealing
temperature of the EI5-center is consistent with our discussion of the
vacancy-interstitial recombination in 3C-SiC. The observation of only
a single annealing stage for the carbon vacancy in the EPR-experiments is also
consistent with PAS experiments\cite{kawasuso:96} performed in irradiated
6H-SiC. There it was found that the corresponding lifetime component annealed at 500$^{\circ}$C.

Recently, Konovalov \emph{et~al}.,\cite{konovalov:01} Zvanut and
Konovalov,\cite{zvanut:02} and Son \emph{et~al}.\cite{son:02} have studied
EPR-centers in semi-insulating 4H-SiC. They identified the centers ID1
(Refs.~\onlinecite{konovalov:01,zvanut:02}) and EI5 (Ref.~\onlinecite{son:02}).
The latter was already observed in irradiated 4H-SiC. The center ID1 resembles
the EI5 center with respect to its $g$-tensor and HF-values, which suggest a
common identification as a carbon vacancy at the cubic lattice
site.\cite{bockstedte:03a} Both centers posses a higher thermal stability than
the EI5-center in irradiated SiC. For example Konovalov \emph{et~al}. have
found that the center persists up to 850$^{\circ}$C.\cite{konovalov:01} The
material used by Konovalov \emph{et~al}. and Zvanut and Konovalov was high
purity, as-grown (0001) 4H-SiC wafers grown by the seeded sublimation method.
The high temperatures ($>2000\,^{\circ}$C) applied during growth suggest that
intrinsic defects are present in the material close to their equilibrium
abundance. This means that carbon vacancies by far outnumber carbon
interstitials, due to the much lower formation energy of the vacancy as
compared to the interstitial.  Also, the concentration of carbon interstitial
clusters is predicted to be well below the vacancy concentration. Therefore
the excess concentration of carbon vacancies in a material grown at high
temperatures may reach its equilibrium concentration at lower temperature only
by a mechanism based on the vacancy diffusion. The large migration
barrier of the carbon vacancy -- 4.1\,eV for V$_{\mathrm{C}}^{+}$ in 3C-SiC --
 explains the high thermal stability of carbon vacancies in as-grown
semi-insulating samples.
\section{Summary and conclusion}
The microscopic picture of the annealing mechanisms of vacancies and
interstitials has been developed from theoretical investigations
based on ab initio methods within the framework of DFT. We analyzed in detail
the ground state configurations of Frenkel pairs formed by carbon (silicon)
vacancies and interstitials and calculated the recombination barriers. We
investigated the capture (and emission) of carbon interstitials by (from)
carbon interstitial clusters. Various clusters involving up to
four carbon interstitials were considered and the energy needed to emit single
carbon interstitials from these clusters was calculated.

A hierarchy of annealing mechanisms has been derived that places the
vacancy-interstitial recombination and the carbon clustering into context with
diffusion based mechanisms and the mestability of the silicon vacancy.  The
recombination barriers and dissociation energies were compared with the
barriers of interstitial and vacancy migration as well as the transformation
barrier of the silicon vacancy. The relevance of the Fermi-level effect was
demonstrated for the annealing hierarchy of the silicon vacancy and
interstitial. The effect originates from two distinct configurations of the
silicon interstitial in p-type and compensated (or n-type) material and the
stabilization of the silicon vacancy in n-type material. In both cases a
strong variation of the migration and transformation barriers with the
Fermi-level is observed, which changes the hierarchical ordering of the
annealing mechanisms. While in compensated material the vacancy-interstitial
recombination and the diffusion-based annealing of silicon interstitials are
activated at lower temperatures than the metastability-related annealing of
the silicon vacancy, this ordering is reversed in p-type material. In n-type
material the metastability-based annealing is unavailable and a diffusion-based
mechanism is observed instead.  Regarding the annealing of carbon
interstitials and vacancies, the highly mobile carbon interstitials were shown
to drive the vacancy-interstitial recombination and the formation of thermally
stable carbon-interstitial clusters at the early stage of the annealing. The
comparably low mobility of the carbon vacancy permits its diffusion-based
annealing only at elevated temperatures. In this case, the hierarchical
ordering was found to be independent of the Fermi-level effect.  The emission
of carbon interstitials from carbon interstitial-clusters is associated with
activation energies that exceed the activation energies of the vacancy
migration. Thus the clusters can provide carbon interstitials even at temperatures
at which isolated carbon vacancies have vanished.

The annealing hierarchy was discussed in the light of annealing experiments on
EPR- and PAS-centers, supported by the recent microscopic identification of
these centers through experimental and theoretical analysis.  A consistent
interpretation of the present experimental data for 3C-SiC is facilitated by
the proposed annealing hierarchy. The available experiments and theoretical
work for 4H- and 6H-SiC suggest that our qualitative conclusions should (with
limitations) apply to other polytypes. We hope to stimulate further
experimental investigation of the defects' annealing kinetics with our
analysis. In particular, the annealing stages of the silicon vacancy above
1000$^{\circ}$C and the confirmation of the tentative assignment of the
EPR-centers T5, EI1 and EI2 to carbon interstitials call for further
clarification. Also the role of defect aggregates, as exemplified by
carbon interstitials, needs experimental substantiation. These
aggregates should provide the missing link between the kinetics of the primary
defects and the thermally stable PL-centers.

\begin{acknowledgments}
Fruitful discussions with G. Pensl, W.J. Choyke, M.E. Zvanut and N.T. Son are
gratefully acknowledged. This work has been supported by the Deutsche
Forschungsgemeinschaft within the SiC Research Group.
\end{acknowledgments}


\begin{thebibliography}{70}
\expandafter\ifx\csname natexlab\endcsname\relax\def\natexlab#1{#1}\fi
\expandafter\ifx\csname bibnamefont\endcsname\relax
  \def\bibnamefont#1{#1}\fi
\expandafter\ifx\csname bibfnamefont\endcsname\relax
  \def\bibfnamefont#1{#1}\fi
\expandafter\ifx\csname citenamefont\endcsname\relax
  \def\citenamefont#1{#1}\fi
\expandafter\ifx\csname url\endcsname\relax
  \def\url#1{\texttt{#1}}\fi
\expandafter\ifx\csname urlprefix\endcsname\relax\def\urlprefix{URL }\fi
\providecommand{\bibinfo}[2]{#2}
\providecommand{\eprint}[2][]{\url{#2}}

\bibitem[{\citenamefont{Choyke and Patrick}(1971)}]{choyke:71}
\bibinfo{author}{\bibfnamefont{W.~J.} \bibnamefont{Choyke}} \bibnamefont{and}
  \bibinfo{author}{\bibfnamefont{L.}~\bibnamefont{Patrick}},
  \bibinfo{journal}{Phys.~Rev.~B} \textbf{\bibinfo{volume}{4}},
  \bibinfo{pages}{1843} (\bibinfo{year}{1971}).

\bibitem[{\citenamefont{Patrick and Choyke}(1972)}]{patrick:72a}
\bibinfo{author}{\bibfnamefont{L.}~\bibnamefont{Patrick}} \bibnamefont{and}
  \bibinfo{author}{\bibfnamefont{W.~J.} \bibnamefont{Choyke}},
  \bibinfo{journal}{Phys.~Rev.~B} \textbf{\bibinfo{volume}{5}},
  \bibinfo{pages}{3253} (\bibinfo{year}{1972}).

\bibitem[{\citenamefont{Egilsson
  et~al.}(1999{\natexlab{a}})\citenamefont{Egilsson, Bergman, Ivanov, Henry,
  and Janz{\'{e}}n}}]{egilsson:99}
\bibinfo{author}{\bibfnamefont{T.}~\bibnamefont{Egilsson}},
  \bibinfo{author}{\bibfnamefont{J.~P.} \bibnamefont{Bergman}},
  \bibinfo{author}{\bibfnamefont{I.~G.} \bibnamefont{Ivanov}},
  \bibinfo{author}{\bibfnamefont{A.}~\bibnamefont{Henry}}, \bibnamefont{and}
  \bibinfo{author}{\bibfnamefont{E.}~\bibnamefont{Janz{\'{e}}n}},
  \bibinfo{journal}{Phys.~Rev.~B} \textbf{\bibinfo{volume}{59}},
  \bibinfo{pages}{1956} (\bibinfo{year}{1999}{\natexlab{a}}).

\bibitem[{\citenamefont{Patrick and Choyke}(1973)}]{patrick:73}
\bibinfo{author}{\bibfnamefont{L.}~\bibnamefont{Patrick}} \bibnamefont{and}
  \bibinfo{author}{\bibfnamefont{W.~J.} \bibnamefont{Choyke}},
  \bibinfo{journal}{J.~Chem.~Phys.~Solids} \textbf{\bibinfo{volume}{34}},
  \bibinfo{pages}{565} (\bibinfo{year}{1973}).

\bibitem[{\citenamefont{Sridhara et~al.}(1998)\citenamefont{Sridhara, Nizhner,
  Devaty, Choyke, Dalibor, Pensl, and Kimoto}}]{sridhara:98a}
\bibinfo{author}{\bibfnamefont{S.~G.} \bibnamefont{Sridhara}},
  \bibinfo{author}{\bibfnamefont{D.~G.} \bibnamefont{Nizhner}},
  \bibinfo{author}{\bibfnamefont{R.~P.} \bibnamefont{Devaty}},
  \bibinfo{author}{\bibfnamefont{W.~J.} \bibnamefont{Choyke}},
  \bibinfo{author}{\bibfnamefont{T.}~\bibnamefont{Dalibor}},
  \bibinfo{author}{\bibfnamefont{G.}~\bibnamefont{Pensl}}, \bibnamefont{and}
  \bibinfo{author}{\bibfnamefont{T.}~\bibnamefont{Kimoto}},
  \bibinfo{journal}{Mater.~Sci.~Forum} \textbf{\bibinfo{volume}{264-268}},
  \bibinfo{pages}{495} (\bibinfo{year}{1998}).

\bibitem[{\citenamefont{Kawasuso et~al.}(2001)\citenamefont{Kawasuso, Redmann,
  Krause-Rehberg, Frank, Weidner, Pensl, Sperr, and Itoh}}]{kawasuso:01}
\bibinfo{author}{\bibfnamefont{A.}~\bibnamefont{Kawasuso}},
  \bibinfo{author}{\bibfnamefont{F.}~\bibnamefont{Redmann}},
  \bibinfo{author}{\bibfnamefont{R.}~\bibnamefont{Krause-Rehberg}},
  \bibinfo{author}{\bibfnamefont{T.}~\bibnamefont{Frank}},
  \bibinfo{author}{\bibfnamefont{M.}~\bibnamefont{Weidner}},
  \bibinfo{author}{\bibfnamefont{G.}~\bibnamefont{Pensl}},
  \bibinfo{author}{\bibfnamefont{P.}~\bibnamefont{Sperr}}, \bibnamefont{and}
  \bibinfo{author}{\bibfnamefont{H.}~\bibnamefont{Itoh}},
  \bibinfo{journal}{J.~Appl.~Phys.} \textbf{\bibinfo{volume}{90}},
  \bibinfo{pages}{3377} (\bibinfo{year}{2001}).

\bibitem[{\citenamefont{Evans et~al.}(2002)\citenamefont{Evans, Steeds, Ley,
  Hundhausen, Schulze, and Pensl}}]{evans:02}
\bibinfo{author}{\bibfnamefont{G.~A.} \bibnamefont{Evans}},
  \bibinfo{author}{\bibfnamefont{J.~W.} \bibnamefont{Steeds}},
  \bibinfo{author}{\bibfnamefont{L.}~\bibnamefont{Ley}},
  \bibinfo{author}{\bibfnamefont{M.}~\bibnamefont{Hundhausen}},
  \bibinfo{author}{\bibfnamefont{N.}~\bibnamefont{Schulze}}, \bibnamefont{and}
  \bibinfo{author}{\bibfnamefont{G.}~\bibnamefont{Pensl}},
  \bibinfo{journal}{Phys.~Rev.~B} \textbf{\bibinfo{volume}{66}},
  \bibinfo{pages}{35204} (\bibinfo{year}{2002}).

\bibitem[{\citenamefont{Freitas et~al.}(1987)\citenamefont{Freitas, Bishop,
  Edmond, Ryu, and Davis}}]{freitas:87}
\bibinfo{author}{\bibfnamefont{J.~A.} \bibnamefont{Freitas}},
  \bibinfo{author}{\bibfnamefont{G.}~\bibnamefont{Bishop}},
  \bibinfo{author}{\bibfnamefont{J.~A.} \bibnamefont{Edmond}},
  \bibinfo{author}{\bibfnamefont{J.}~\bibnamefont{Ryu}}, \bibnamefont{and}
  \bibinfo{author}{\bibfnamefont{R.~F.} \bibnamefont{Davis}},
  \bibinfo{journal}{J.~Appl.~Phys.} \textbf{\bibinfo{volume}{61}},
  \bibinfo{pages}{2011} (\bibinfo{year}{1987}).

\bibitem[{\citenamefont{Egilsson
  et~al.}(1999{\natexlab{b}})\citenamefont{Egilsson, Henry, Ivanov,
  Lindstr{\"{o}}m, and Janz{\'{e}}n}}]{egilsson:99a}
\bibinfo{author}{\bibfnamefont{T.}~\bibnamefont{Egilsson}},
  \bibinfo{author}{\bibfnamefont{A.}~\bibnamefont{Henry}},
  \bibinfo{author}{\bibfnamefont{I.~G.} \bibnamefont{Ivanov}},
  \bibinfo{author}{\bibfnamefont{J.~L.} \bibnamefont{Lindstr{\"{o}}m}},
  \bibnamefont{and}
  \bibinfo{author}{\bibfnamefont{E.}~\bibnamefont{Janz{\'{e}}n}},
  \bibinfo{journal}{Phys.~Rev.~B} \textbf{\bibinfo{volume}{59}},
  \bibinfo{pages}{8008} (\bibinfo{year}{1999}{\natexlab{b}}).

\bibitem[{\citenamefont{Kawasuso et~al.}(1996)\citenamefont{Kawasuso, Itoh,
  Okada, and Okumura}}]{kawasuso:96}
\bibinfo{author}{\bibfnamefont{A.}~\bibnamefont{Kawasuso}},
  \bibinfo{author}{\bibfnamefont{H.}~\bibnamefont{Itoh}},
  \bibinfo{author}{\bibfnamefont{S.}~\bibnamefont{Okada}}, \bibnamefont{and}
  \bibinfo{author}{\bibfnamefont{H.}~\bibnamefont{Okumura}},
  \bibinfo{journal}{J.~Appl.~Phys.} \textbf{\bibinfo{volume}{80}},
  \bibinfo{pages}{5639} (\bibinfo{year}{1996}).

\bibitem[{\citenamefont{Brauer et~al.}(1996{\natexlab{a}})\citenamefont{Brauer,
  Anwand, Coleman, Knights, Plazaola, Pacaud, Skorupa, St{\"o}rmer, and
  Willutzki}}]{brauer:96}
\bibinfo{author}{\bibfnamefont{G.}~\bibnamefont{Brauer}},
  \bibinfo{author}{\bibfnamefont{W.}~\bibnamefont{Anwand}},
  \bibinfo{author}{\bibfnamefont{P.~G.} \bibnamefont{Coleman}},
  \bibinfo{author}{\bibfnamefont{A.~P.} \bibnamefont{Knights}},
  \bibinfo{author}{\bibfnamefont{F.}~\bibnamefont{Plazaola}},
  \bibinfo{author}{\bibfnamefont{Y.}~\bibnamefont{Pacaud}},
  \bibinfo{author}{\bibfnamefont{W.}~\bibnamefont{Skorupa}},
  \bibinfo{author}{\bibfnamefont{J.}~\bibnamefont{St{\"o}rmer}},
  \bibnamefont{and}
  \bibinfo{author}{\bibfnamefont{P.}~\bibnamefont{Willutzki}},
  \bibinfo{journal}{Phys.~Rev.~B} \textbf{\bibinfo{volume}{54}},
  \bibinfo{pages}{3084} (\bibinfo{year}{1996}{\natexlab{a}}).

\bibitem[{\citenamefont{Kawasuso et~al.}(1998)\citenamefont{Kawasuso, Itoh,
  Morishita, Yoshikawa, and Ohshima}}]{kawasuso:98}
\bibinfo{author}{\bibfnamefont{A.}~\bibnamefont{Kawasuso}},
  \bibinfo{author}{\bibfnamefont{H.}~\bibnamefont{Itoh}},
  \bibinfo{author}{\bibfnamefont{N.}~\bibnamefont{Morishita}},
  \bibinfo{author}{\bibfnamefont{M.}~\bibnamefont{Yoshikawa}},
  \bibnamefont{and} \bibinfo{author}{\bibfnamefont{T.}~\bibnamefont{Ohshima}},
  \bibinfo{journal}{Appl.~Phys.~A} \textbf{\bibinfo{volume}{67}},
  \bibinfo{pages}{209} (\bibinfo{year}{1998}).

\bibitem[{\citenamefont{Polity et~al.}(1999)\citenamefont{Polity, Huth, and
  Lausmann}}]{polity:99}
\bibinfo{author}{\bibfnamefont{A.}~\bibnamefont{Polity}},
  \bibinfo{author}{\bibfnamefont{S.}~\bibnamefont{Huth}}, \bibnamefont{and}
  \bibinfo{author}{\bibfnamefont{M.}~\bibnamefont{Lausmann}},
  \bibinfo{journal}{Phys.~Rev.~B} \textbf{\bibinfo{volume}{59}},
  \bibinfo{pages}{10603} (\bibinfo{year}{1999}).

\bibitem[{\citenamefont{Wimbauer et~al.}(1997)\citenamefont{Wimbauer, Meyer,
  Hofstaetter, Scharmann, and Overhof}}]{wimbauer:97}
\bibinfo{author}{\bibfnamefont{T.}~\bibnamefont{Wimbauer}},
  \bibinfo{author}{\bibfnamefont{B.~K.} \bibnamefont{Meyer}},
  \bibinfo{author}{\bibfnamefont{A.}~\bibnamefont{Hofstaetter}},
  \bibinfo{author}{\bibfnamefont{A.}~\bibnamefont{Scharmann}},
  \bibnamefont{and} \bibinfo{author}{\bibfnamefont{H.}~\bibnamefont{Overhof}},
  \bibinfo{journal}{Phys.~Rev.~B} \textbf{\bibinfo{volume}{56}},
  \bibinfo{pages}{7384} (\bibinfo{year}{1997}).

\bibitem[{\citenamefont{von Bardeleben et~al.}(2000)\citenamefont{von
  Bardeleben, Cantin, Henry, and Barthe}}]{vonbardeleben:00}
\bibinfo{author}{\bibfnamefont{H.~J.} \bibnamefont{von Bardeleben}},
  \bibinfo{author}{\bibfnamefont{J.~L.} \bibnamefont{Cantin}},
  \bibinfo{author}{\bibfnamefont{L.}~\bibnamefont{Henry}}, \bibnamefont{and}
  \bibinfo{author}{\bibfnamefont{M.~F.} \bibnamefont{Barthe}},
  \bibinfo{journal}{Phys.~Rev.~B} \textbf{\bibinfo{volume}{62}},
  \bibinfo{pages}{10841} (\bibinfo{year}{2000}).

\bibitem[{\citenamefont{Son et~al.}(2001{\natexlab{a}})\citenamefont{Son, Hai,
  and Janz{\'{e}}n}}]{son:01}
\bibinfo{author}{\bibfnamefont{N.~T.} \bibnamefont{Son}},
  \bibinfo{author}{\bibfnamefont{P.~N.} \bibnamefont{Hai}}, \bibnamefont{and}
  \bibinfo{author}{\bibfnamefont{E.}~\bibnamefont{Janz{\'{e}}n}},
  \bibinfo{journal}{Phys.~Rev.~B} \textbf{\bibinfo{volume}{63}},
  \bibinfo{pages}{201201} (\bibinfo{year}{2001}{\natexlab{a}}).

\bibitem[{\citenamefont{Itoh et~al.}(1997)\citenamefont{Itoh, Kawasuso,
  Ohshima, Yoshikawa, Nashiyama, Tanigawa, Misawa, Okumura, and
  Yoshida}}]{itoh:97a}
\bibinfo{author}{\bibfnamefont{H.}~\bibnamefont{Itoh}},
  \bibinfo{author}{\bibfnamefont{A.}~\bibnamefont{Kawasuso}},
  \bibinfo{author}{\bibfnamefont{T.}~\bibnamefont{Ohshima}},
  \bibinfo{author}{\bibfnamefont{M.}~\bibnamefont{Yoshikawa}},
  \bibinfo{author}{\bibfnamefont{I.}~\bibnamefont{Nashiyama}},
  \bibinfo{author}{\bibfnamefont{S.}~\bibnamefont{Tanigawa}},
  \bibinfo{author}{\bibfnamefont{S.}~\bibnamefont{Misawa}},
  \bibinfo{author}{\bibfnamefont{H.}~\bibnamefont{Okumura}}, \bibnamefont{and}
  \bibinfo{author}{\bibfnamefont{S.}~\bibnamefont{Yoshida}},
  \bibinfo{journal}{phys.~stat.~sol.~({a})} \textbf{\bibinfo{volume}{162}},
  \bibinfo{pages}{173} (\bibinfo{year}{1997}).

\bibitem[{\citenamefont{Brauer et~al.}(1996{\natexlab{b}})\citenamefont{Brauer,
  Anwand, Nicht, Kuriplach, Sob, Wagner, Coleman, Puska, and
  Korhonen}}]{brauer:96a}
\bibinfo{author}{\bibfnamefont{G.}~\bibnamefont{Brauer}},
  \bibinfo{author}{\bibfnamefont{W.}~\bibnamefont{Anwand}},
  \bibinfo{author}{\bibfnamefont{E.-M.} \bibnamefont{Nicht}},
  \bibinfo{author}{\bibfnamefont{J.}~\bibnamefont{Kuriplach}},
  \bibinfo{author}{\bibfnamefont{M.}~\bibnamefont{Sob}},
  \bibinfo{author}{\bibfnamefont{N.}~\bibnamefont{Wagner}},
  \bibinfo{author}{\bibfnamefont{P.~G.} \bibnamefont{Coleman}},
  \bibinfo{author}{\bibfnamefont{M.~J.} \bibnamefont{Puska}}, \bibnamefont{and}
  \bibinfo{author}{\bibfnamefont{T.}~\bibnamefont{Korhonen}},
  \bibinfo{journal}{Phys.~Rev.~B} \textbf{\bibinfo{volume}{54}},
  \bibinfo{pages}{2512} (\bibinfo{year}{1996}{\natexlab{b}}).

\bibitem[{\citenamefont{Staab et~al.}(2001)\citenamefont{Staab, Torpo, Puska,
  and Nieminen}}]{staab:01}
\bibinfo{author}{\bibfnamefont{T.~E.~M.} \bibnamefont{Staab}},
  \bibinfo{author}{\bibfnamefont{L.~M.} \bibnamefont{Torpo}},
  \bibinfo{author}{\bibfnamefont{M.~J.} \bibnamefont{Puska}}, \bibnamefont{and}
  \bibinfo{author}{\bibfnamefont{R.~M.} \bibnamefont{Nieminen}},
  \bibinfo{journal}{Mater.~Sci.~Forum} \textbf{\bibinfo{volume}{353-356}},
  \bibinfo{pages}{533} (\bibinfo{year}{2001}).

\bibitem[{\citenamefont{Konovalov et~al.}(2001)\citenamefont{Konovalov, Zvanut,
  Tsvetkov, Jenny, M{\"u}ller, and Hobsgood}}]{konovalov:01}
\bibinfo{author}{\bibfnamefont{V.~V.} \bibnamefont{Konovalov}},
  \bibinfo{author}{\bibfnamefont{M.~E.} \bibnamefont{Zvanut}},
  \bibinfo{author}{\bibfnamefont{V.~F.} \bibnamefont{Tsvetkov}},
  \bibinfo{author}{\bibfnamefont{J.~R.} \bibnamefont{Jenny}},
  \bibinfo{author}{\bibfnamefont{S.~G.} \bibnamefont{M{\"u}ller}},
  \bibnamefont{and} \bibinfo{author}{\bibfnamefont{H.~M.}
  \bibnamefont{Hobsgood}}, \bibinfo{journal}{Physica B}
  \textbf{\bibinfo{volume}{308-310}}, \bibinfo{pages}{671}
  (\bibinfo{year}{2001}).

\bibitem[{\citenamefont{Bratus et~al.}(2001)\citenamefont{Bratus, Makeeva,
  Okulov, Petrenko, Petrenko, and von Bardeleben}}]{bratus:01}
\bibinfo{author}{\bibfnamefont{V.~Y.} \bibnamefont{Bratus}},
  \bibinfo{author}{\bibfnamefont{I.~N.} \bibnamefont{Makeeva}},
  \bibinfo{author}{\bibfnamefont{S.~M.} \bibnamefont{Okulov}},
  \bibinfo{author}{\bibfnamefont{T.~L.} \bibnamefont{Petrenko}},
  \bibinfo{author}{\bibfnamefont{T.~T.} \bibnamefont{Petrenko}},
  \bibnamefont{and} \bibinfo{author}{\bibfnamefont{H.}~\bibnamefont{von
  Bardeleben}}, \bibinfo{journal}{Physica B}
  \textbf{\bibinfo{volume}{309-310}}, \bibinfo{pages}{621}
  (\bibinfo{year}{2001}).

\bibitem[{\citenamefont{Petrenko et~al.}(2001)\citenamefont{Petrenko, Petrenko,
  Bratus, and Monge}}]{petrenko:01}
\bibinfo{author}{\bibfnamefont{T.~T.} \bibnamefont{Petrenko}},
  \bibinfo{author}{\bibfnamefont{T.~L.} \bibnamefont{Petrenko}},
  \bibinfo{author}{\bibfnamefont{V.~Y.} \bibnamefont{Bratus}},
  \bibnamefont{and} \bibinfo{author}{\bibfnamefont{J.~L.} \bibnamefont{Monge}},
  \bibinfo{journal}{Physica B} \textbf{\bibinfo{volume}{308-310}},
  \bibinfo{pages}{637} (\bibinfo{year}{2001}).

\bibitem[{\citenamefont{Bockstedte et~al.}(2002)\citenamefont{Bockstedte, Heid,
  Mattausch, and Pankratov}}]{bockstedte:02}
\bibinfo{author}{\bibfnamefont{M.}~\bibnamefont{Bockstedte}},
  \bibinfo{author}{\bibfnamefont{M.}~\bibnamefont{Heid}},
  \bibinfo{author}{\bibfnamefont{A.}~\bibnamefont{Mattausch}},
  \bibnamefont{and}
  \bibinfo{author}{\bibfnamefont{O.}~\bibnamefont{Pankratov}},
  \bibinfo{journal}{Mater.~Sci.~Forum} \textbf{\bibinfo{volume}{389-393}},
  \bibinfo{pages}{471} (\bibinfo{year}{2002}).

\bibitem[{\citenamefont{Bockstedte
  et~al.}(2003{\natexlab{a}})\citenamefont{Bockstedte, Heid, and
  Pankratov}}]{bockstedte:03a}
\bibinfo{author}{\bibfnamefont{M.}~\bibnamefont{Bockstedte}},
  \bibinfo{author}{\bibfnamefont{M.}~\bibnamefont{Heid}}, \bibnamefont{and}
  \bibinfo{author}{\bibfnamefont{O.}~\bibnamefont{Pankratov}},
  \bibinfo{journal}{Phys.~Rev.~B} \textbf{\bibinfo{volume}{67}},
  \bibinfo{pages}{193102} (\bibinfo{year}{2003}{\natexlab{a}}).

\bibitem[{\citenamefont{Kawasuso et~al.}(1997)\citenamefont{Kawasuso, Itoh,
  Ohshima, Abe, and Okada}}]{kawasuso:97}
\bibinfo{author}{\bibfnamefont{A.}~\bibnamefont{Kawasuso}},
  \bibinfo{author}{\bibfnamefont{H.}~\bibnamefont{Itoh}},
  \bibinfo{author}{\bibfnamefont{T.}~\bibnamefont{Ohshima}},
  \bibinfo{author}{\bibfnamefont{K.}~\bibnamefont{Abe}}, \bibnamefont{and}
  \bibinfo{author}{\bibfnamefont{S.}~\bibnamefont{Okada}},
  \bibinfo{journal}{J.~Appl.~Phys.} \textbf{\bibinfo{volume}{82}},
  \bibinfo{pages}{3232} (\bibinfo{year}{1997}).

\bibitem[{\citenamefont{Lingner et~al.}(2001)\citenamefont{Lingner,
  Greulich-Weber, Spaeth, Gerstmann, Rauls, Hajnal, Frauenheim, and
  Overhof}}]{lingner:01}
\bibinfo{author}{\bibfnamefont{T.}~\bibnamefont{Lingner}},
  \bibinfo{author}{\bibfnamefont{S.}~\bibnamefont{Greulich-Weber}},
  \bibinfo{author}{\bibfnamefont{J.-M.} \bibnamefont{Spaeth}},
  \bibinfo{author}{\bibfnamefont{U.}~\bibnamefont{Gerstmann}},
  \bibinfo{author}{\bibfnamefont{E.}~\bibnamefont{Rauls}},
  \bibinfo{author}{\bibfnamefont{Z.}~\bibnamefont{Hajnal}},
  \bibinfo{author}{\bibfnamefont{T.}~\bibnamefont{Frauenheim}},
  \bibnamefont{and} \bibinfo{author}{\bibfnamefont{H.}~\bibnamefont{Overhof}},
  \bibinfo{journal}{Phys.~Rev.~B} \textbf{\bibinfo{volume}{64}},
  \bibinfo{pages}{245212} (\bibinfo{year}{2001}).

\bibitem[{\citenamefont{Son et~al.}(1997)\citenamefont{Son, S{\"o}rman, Chen,
  Hallin, Kordina, Monemar, and Janz{\'e}n}}]{son:97}
\bibinfo{author}{\bibfnamefont{N.~T.} \bibnamefont{Son}},
  \bibinfo{author}{\bibfnamefont{E.}~\bibnamefont{S{\"{o}}rman}},
  \bibinfo{author}{\bibfnamefont{W.~M.} \bibnamefont{Chen}},
  \bibinfo{author}{\bibfnamefont{C.}~\bibnamefont{Hallin}},
  \bibinfo{author}{\bibfnamefont{O.}~\bibnamefont{Kordina}},
  \bibinfo{author}{\bibfnamefont{B.}~\bibnamefont{Monemar}},
  \bibinfo{author}{\bibfnamefont{E.}~\bibnamefont{Janz{\'e}n},
  \bibnamefont{and}
  \bibinfo{author}{\bibfnamefont{J.~L.} \bibnamefont{Lindstr{\"{o}}m}},},
  \bibinfo{journal}{Phys.~Rev.~B} \textbf{\bibinfo{volume}{55}},
  \bibinfo{pages}{2863} (\bibinfo{year}{1997}).

\bibitem[{\citenamefont{Rauls et~al.}(2000)\citenamefont{Rauls, Lingner,
  Hajnal, Greulich-Weber, Frauenheim, and Spaeth}}]{rauls:00}
\bibinfo{author}{\bibfnamefont{E.}~\bibnamefont{Rauls}},
  \bibinfo{author}{\bibfnamefont{T.}~\bibnamefont{Lingner}},
  \bibinfo{author}{\bibfnamefont{Z.}~\bibnamefont{Hajnal}},
  \bibinfo{author}{\bibfnamefont{S.}~\bibnamefont{Greulich-Weber}},
  \bibinfo{author}{\bibfnamefont{T.}~\bibnamefont{Frauenheim}},
  \bibnamefont{and} \bibinfo{author}{\bibfnamefont{J.-M.}
  \bibnamefont{Spaeth}}, \bibinfo{journal}{phys.~stat.~sol.~({b})}
  \textbf{\bibinfo{volume}{217}}, \bibinfo{pages}{R1} (\bibinfo{year}{2000}).

\bibitem[{\citenamefont{Bockstedte and Pankratov}(2000)}]{bockstedte:00}
\bibinfo{author}{\bibfnamefont{M.}~\bibnamefont{Bockstedte}} \bibnamefont{and}
  \bibinfo{author}{\bibfnamefont{O.}~\bibnamefont{Pankratov}},
  \bibinfo{journal}{Mater.~Sci.~Forum} \textbf{\bibinfo{volume}{338-342}},
  \bibinfo{pages}{949} (\bibinfo{year}{2000}).

\bibitem[{\citenamefont{Son et~al.}(2003)\citenamefont{Son, Magnusson, Zolnai,
  Ellison, and Janz{\'e}n}}]{son:03}
\bibinfo{author}{\bibfnamefont{N.}~\bibnamefont{Son}},
  \bibinfo{author}{\bibfnamefont{B.}~\bibnamefont{Magnusson}},
  \bibinfo{author}{\bibfnamefont{Z.}~\bibnamefont{Zolnai}},
  \bibinfo{author}{\bibfnamefont{A.}~\bibnamefont{Ellison}}, \bibnamefont{and}
  \bibinfo{author}{\bibfnamefont{E.}~\bibnamefont{Janz{\'e}n}},
  \bibinfo{journal}{Mater.~Sci.~Forum} \textbf{\bibinfo{volume}{433-436}},
  \bibinfo{pages}{45} (\bibinfo{year}{2003}).

\bibitem[{\citenamefont{Bockstedte et~al.}(1997)\citenamefont{Bockstedte, Kley,
  Neugebauer, and Scheffler}}]{bockstedte:97a}
\bibinfo{author}{\bibfnamefont{M.}~\bibnamefont{Bockstedte}},
  \bibinfo{author}{\bibfnamefont{A.}~\bibnamefont{Kley}},
  \bibinfo{author}{\bibfnamefont{J.}~\bibnamefont{Neugebauer}},
  \bibnamefont{and}
  \bibinfo{author}{\bibfnamefont{M.}~\bibnamefont{Scheffler}},
  \bibinfo{journal}{Comp.~Phys.~Comm.} \textbf{\bibinfo{volume}{200}},
  \bibinfo{pages}{107} (\bibinfo{year}{1997}).

\bibitem[{\citenamefont{Hohenberg and Kohn}(1964)}]{hohenberg:64}
\bibinfo{author}{\bibfnamefont{P.}~\bibnamefont{Hohenberg}} \bibnamefont{and}
  \bibinfo{author}{\bibfnamefont{W.}~\bibnamefont{Kohn}},
  \bibinfo{journal}{Phys.~Rev.} \textbf{\bibinfo{volume}{136B}},
  \bibinfo{pages}{864} (\bibinfo{year}{1964}).

\bibitem[{\citenamefont{Kohn and Sham}(1965)}]{kohn:65}
\bibinfo{author}{\bibfnamefont{W.}~\bibnamefont{Kohn}} \bibnamefont{and}
  \bibinfo{author}{\bibfnamefont{J.~L.} \bibnamefont{Sham}},
  \bibinfo{journal}{Phys.~Rev.} \textbf{\bibinfo{volume}{140A}},
  \bibinfo{pages}{1133} (\bibinfo{year}{1965}).

\bibitem[{\citenamefont{Perdew and Zunger}(1981)}]{perdew:81}
\bibinfo{author}{\bibfnamefont{J.~P.} \bibnamefont{Perdew}} \bibnamefont{and}
  \bibinfo{author}{\bibfnamefont{A.}~\bibnamefont{Zunger}},
  \bibinfo{journal}{Phys.~Rev.~B} \textbf{\bibinfo{volume}{23}},
  \bibinfo{pages}{5048} (\bibinfo{year}{1981}).

\bibitem[{\citenamefont{Ceperley and Alder}(1980)}]{ceperley:80}
\bibinfo{author}{\bibfnamefont{D.~M.} \bibnamefont{Ceperley}} \bibnamefont{and}
  \bibinfo{author}{\bibfnamefont{B.~J.} \bibnamefont{Alder}},
  \bibinfo{journal}{Phys.~Rev.~Lett.} \textbf{\bibinfo{volume}{45}},
  \bibinfo{pages}{566} (\bibinfo{year}{1980}).

\bibitem[{\citenamefont{Monkhorst and Pack}(1976)}]{monkhorst:76}
\bibinfo{author}{\bibfnamefont{H.~J.} \bibnamefont{Monkhorst}}
  \bibnamefont{and} \bibinfo{author}{\bibfnamefont{J.~D.} \bibnamefont{Pack}},
  \bibinfo{journal}{Phys.~Rev.~B} \textbf{\bibinfo{volume}{13}},
  \bibinfo{pages}{5188} (\bibinfo{year}{1976}).

\bibitem[{\citenamefont{Makov and Payne}(1995)}]{makov:95}
\bibinfo{author}{\bibfnamefont{G.}~\bibnamefont{Makov}} \bibnamefont{and}
  \bibinfo{author}{\bibfnamefont{M.~C.} \bibnamefont{Payne}},
  \bibinfo{journal}{Phys.~Rev.~B} \textbf{\bibinfo{volume}{51}},
  \bibinfo{pages}{4014} (\bibinfo{year}{1995}).

\bibitem[{\citenamefont{Lento et~al.}(2002)\citenamefont{Lento, Mozos, and
  Nieminen}}]{lento:02}
\bibinfo{author}{\bibfnamefont{J.}~\bibnamefont{Lento}},
  \bibinfo{author}{\bibfnamefont{J.-L.} \bibnamefont{Mozos}}, \bibnamefont{and}
  \bibinfo{author}{\bibfnamefont{R.~M.} \bibnamefont{Nieminen}},
  \bibinfo{journal}{J.~Phys.~Cond.~Matt.} \textbf{\bibinfo{volume}{14}},
  \bibinfo{pages}{2637} (\bibinfo{year}{2002}).

\bibitem[{\citenamefont{Bockstedte
  et~al.}(2003{\natexlab{b}})\citenamefont{Bockstedte, Mattausch, and
  Pankratov}}]{bockstedte:03b}
\bibinfo{author}{\bibfnamefont{M.}~\bibnamefont{Bockstedte}},
  \bibinfo{author}{\bibfnamefont{A.}~\bibnamefont{Mattausch}},
  \bibnamefont{and}
  \bibinfo{author}{\bibfnamefont{O.}~\bibnamefont{Pankratov}},
  \bibinfo{journal}{Phys.~Rev.~B} {\bibinfo{volume}{in print}}.

\bibitem[{\citenamefont{Troullier and Martins}(1991)}]{troullier:91}
\bibinfo{author}{\bibfnamefont{N.}~\bibnamefont{Troullier}} \bibnamefont{and}
  \bibinfo{author}{\bibfnamefont{J.~L.} \bibnamefont{Martins}},
  \bibinfo{journal}{Phys.~Rev.~B} \textbf{\bibinfo{volume}{43}},
  \bibinfo{pages}{1993} (\bibinfo{year}{1991}).

\bibitem[{\citenamefont{Fuchs and Scheffler}(1999)}]{fuchs:99}
\bibinfo{author}{\bibfnamefont{M.}~\bibnamefont{Fuchs}} \bibnamefont{and}
  \bibinfo{author}{\bibfnamefont{M.}~\bibnamefont{Scheffler}},
  \bibinfo{journal}{Comp.~Phys.~Comm.} \textbf{\bibinfo{volume}{119}},
  \bibinfo{pages}{67} (\bibinfo{year}{1999}).
\bibitem{pseudo}
  A carbon pseudopotential with the following matching radii is used: $r^{{\rm
      c}}_{{\rm s}}$=1.6\,Bohr, $r^{{\rm c}}_{{\rm p}}$=1.7\,Bohr and $r^{{\rm
      c}}_{{\rm d}}$=1.5\,Bohr.
\bibitem[{\citenamefont{Baraff et~al.}(1979)\citenamefont{Baraff, Kane, and
  Schl{\"u}ter}}]{baraff:79}
\bibinfo{author}{\bibfnamefont{G.~A.} \bibnamefont{Baraff}},
  \bibinfo{author}{\bibfnamefont{E.O.}~\bibnamefont{Kane}}, \bibnamefont{and}
  \bibinfo{author}{\bibfnamefont{M.}~\bibnamefont{Schl{\"u}ter}},
  \bibinfo{journal}{Phys.~Rev.~Lett.} \textbf{\bibinfo{volume}{43}},
  \bibinfo{pages}{956} (\bibinfo{year}{1979}).

\bibitem[{\citenamefont{Pehlke and Kratzer}(2000)}]{pehlke:00p}
\bibinfo{author}{\bibfnamefont{E.}~\bibnamefont{Pehlke}} \bibnamefont{and}
  \bibinfo{author}{\bibfnamefont{P.}~\bibnamefont{Kratzer}}
  (\bibinfo{year}{2000}), \bibinfo{note}{priv. commun.}

\bibitem[{\citenamefont{Ionova and Carter}(1993)}]{ionova:93}
\bibinfo{author}{\bibfnamefont{I.~V.} \bibnamefont{Ionova}} \bibnamefont{and}
  \bibinfo{author}{\bibfnamefont{E.~A.} \bibnamefont{Carter}},
  \bibinfo{journal}{J.~Chem.~Phys.} \textbf{\bibinfo{volume}{98}},
  \bibinfo{pages}{6377} (\bibinfo{year}{1993}).

\bibitem[{\citenamefont{Mattausch
  et~al.}(2001{\natexlab{a}})\citenamefont{Mattausch, Bockstedte, and
  Pankratov}}]{mattausch:01}
\bibinfo{author}{\bibfnamefont{A.}~\bibnamefont{Mattausch}},
  \bibinfo{author}{\bibfnamefont{M.}~\bibnamefont{Bockstedte}},
  \bibnamefont{and}
  \bibinfo{author}{\bibfnamefont{O.}~\bibnamefont{Pankratov}},
  \bibinfo{journal}{Mater.~Sci.~Forum} \textbf{\bibinfo{volume}{353-356}},
  \bibinfo{pages}{323} (\bibinfo{year}{2001}{\natexlab{a}}).

\bibitem[{\citenamefont{Bockstedte
  et~al.}(2003{\natexlab{c}})\citenamefont{Bockstedte, Heid, Mattausch, and
  Pankratov}}]{bockstedte:03}
\bibinfo{author}{\bibfnamefont{M.}~\bibnamefont{Bockstedte}},
  \bibinfo{author}{\bibfnamefont{M.}~\bibnamefont{Heid}},
  \bibinfo{author}{\bibfnamefont{A.}~\bibnamefont{Mattausch}},
  \bibnamefont{and}
  \bibinfo{author}{\bibfnamefont{O.}~\bibnamefont{Pankratov}},
  \bibinfo{journal}{Mater.~Sci.~Forum} \textbf{\bibinfo{volume}{433-436}},
  \bibinfo{pages}{471} (\bibinfo{year}{2003}{\natexlab{c}}).

\bibitem[{\citenamefont{Zywietz et~al.}(1999)\citenamefont{Zywietz,
  Furthm{\"u}ller, and Bechstedt}}]{zywietz:99}
\bibinfo{author}{\bibfnamefont{A.}~\bibnamefont{Zywietz}},
  \bibinfo{author}{\bibfnamefont{J.}~\bibnamefont{Furthm{\"u}ller}},
  \bibnamefont{and}
  \bibinfo{author}{\bibfnamefont{F.}~\bibnamefont{Bechstedt}},
  \bibinfo{journal}{Phys.~Rev.~B} \textbf{\bibinfo{volume}{59}},
  \bibinfo{pages}{15166} (\bibinfo{year}{1999}).

\bibitem[{\citenamefont{Torpo et~al.}(2001)\citenamefont{Torpo, Marlo, Staab,
  and Nieminen}}]{torpo:01}
\bibinfo{author}{\bibfnamefont{L.}~\bibnamefont{Torpo}},
  \bibinfo{author}{\bibfnamefont{M.}~\bibnamefont{Marlo}},
  \bibinfo{author}{\bibfnamefont{T.~E.~M.} \bibnamefont{Staab}},
  \bibnamefont{and} \bibinfo{author}{\bibfnamefont{R.~M.}
  \bibnamefont{Nieminen}}, \bibinfo{journal}{J.~Phys.~Cond.~Matt.}
  \textbf{\bibinfo{volume}{13}}, \bibinfo{pages}{6203} (\bibinfo{year}{2001}).

\bibitem[{\citenamefont{Rauls et~al.}(2001)\citenamefont{Rauls, Staab, Hajnal,
  and Frauenheim}}]{rauls:01}
\bibinfo{author}{\bibfnamefont{E.}~\bibnamefont{Rauls}},
  \bibinfo{author}{\bibfnamefont{T.~E.~M.} \bibnamefont{Staab}},
  \bibinfo{author}{\bibfnamefont{Z.}~\bibnamefont{Hajnal}}, \bibnamefont{and}
  \bibinfo{author}{\bibfnamefont{T.}~\bibnamefont{Frauenheim}},
  \bibinfo{journal}{Physica B} \textbf{\bibinfo{volume}{308-310}},
  \bibinfo{pages}{645} (\bibinfo{year}{2001}).

\bibitem[{\citenamefont{Torpo et~al.}(2002)\citenamefont{Torpo, Staab, and
  Nieminen}}]{torpo:02}
\bibinfo{author}{\bibfnamefont{L.}~\bibnamefont{Torpo}},
  \bibinfo{author}{\bibfnamefont{T.~E.~M.} \bibnamefont{Staab}},
  \bibnamefont{and} \bibinfo{author}{\bibfnamefont{R.~M.}
  \bibnamefont{Nieminen}}, \bibinfo{journal}{Phys.~Rev.~B}
  \textbf{\bibinfo{volume}{65}}, \bibinfo{pages}{85202} (\bibinfo{year}{2002}).

\bibitem[{\citenamefont{Goss et~al.}(2001)\citenamefont{Goss, Coomer, Jones,
  Shaw, Briddon, Rayson, and {\"{O}}berg}}]{goss:01}
\bibinfo{author}{\bibfnamefont{J.~P.} \bibnamefont{Goss}},
  \bibinfo{author}{\bibfnamefont{B.~J.} \bibnamefont{Coomer}},
  \bibinfo{author}{\bibfnamefont{R.}~\bibnamefont{Jones}},
  \bibinfo{author}{\bibfnamefont{T.~D.} \bibnamefont{Shaw}},
  \bibinfo{author}{\bibfnamefont{P.~R.} \bibnamefont{Briddon}},
  \bibinfo{author}{\bibfnamefont{M.}~\bibnamefont{Rayson}}, \bibnamefont{and}
  \bibinfo{author}{\bibfnamefont{S.}~\bibnamefont{{\"{O}}berg}},
  \bibinfo{journal}{Phys.~Rev.~B} \textbf{\bibinfo{volume}{63}},
  \bibinfo{pages}{195208} (\bibinfo{year}{2001}).

\bibitem[{\citenamefont{Mattausch et~al.}(2003)\citenamefont{Mattausch,
  Bockstedte, and Pankratov}}]{mattausch:03a}
\bibinfo{author}{\bibfnamefont{A.}~\bibnamefont{Mattausch}},
  \bibinfo{author}{\bibfnamefont{M.}~\bibnamefont{Bockstedte}},
  \bibnamefont{and}
  \bibinfo{author}{\bibfnamefont{O.}~\bibnamefont{Pankratov}},
  unpublished.

\bibitem[{\citenamefont{Bourgoin and Lannoo}(1983)}]{bourgoin:83}
\bibinfo{author}{\bibfnamefont{J.}~\bibnamefont{Bourgoin}} \bibnamefont{and}
  \bibinfo{author}{\bibfnamefont{M.}~\bibnamefont{Lannoo}},
  \emph{\bibinfo{title}{Point Defects in Semiconductors {II}}},
  vol.~\bibinfo{volume}{35} of \emph{\bibinfo{series}{Springer Series in
  Solid-State Sciences}} (\bibinfo{publisher}{Springer Verlag},
  \bibinfo{address}{Berlin}, \bibinfo{year}{1983}).

\bibitem[{\citenamefont{Itoh et~al.}(1989)\citenamefont{Itoh, Hayakawa,
  Nashiyama, and Sakuma}}]{itoh:89}
\bibinfo{author}{\bibfnamefont{H.}~\bibnamefont{Itoh}},
  \bibinfo{author}{\bibfnamefont{N.}~\bibnamefont{Hayakawa}},
  \bibinfo{author}{\bibfnamefont{I.}~\bibnamefont{Nashiyama}},
  \bibnamefont{and} \bibinfo{author}{\bibfnamefont{E.}~\bibnamefont{Sakuma}},
  \bibinfo{journal}{J.~Appl.~Phys.} \textbf{\bibinfo{volume}{66}},
  \bibinfo{pages}{4529} (\bibinfo{year}{1989}).

\bibitem[{\citenamefont{Vainer and Il'in}(1981{\natexlab{a}})}]{vainer:81}
\bibinfo{author}{\bibfnamefont{V.~S.} \bibnamefont{Vainer}} \bibnamefont{and}
  \bibinfo{author}{\bibfnamefont{V.~A.} \bibnamefont{Il'in}},
  \bibinfo{journal}{Sov. Phys. Solid. State} \textbf{\bibinfo{volume}{23}},
  \bibinfo{pages}{2126} (\bibinfo{year}{1981}{\natexlab{a}}).

\bibitem[{\citenamefont{Gerstmann et~al.}(2003)\citenamefont{Gerstmann, Rauls,
  Frauenheim, and Overhof}}]{gerstmann:03}
\bibinfo{author}{\bibfnamefont{U.}~\bibnamefont{Gerstmann}},
  \bibinfo{author}{\bibfnamefont{E.}~\bibnamefont{Rauls}},
  \bibinfo{author}{\bibfnamefont{T.}~\bibnamefont{Frauenheim}},
  \bibnamefont{and} \bibinfo{author}{\bibfnamefont{H.}~\bibnamefont{Overhof}},
  \bibinfo{journal}{Phys.~Rev.~B} \textbf{\bibinfo{volume}{67}},
  \bibinfo{pages}{205202} (\bibinfo{year}{2003}).

\bibitem[{\citenamefont{Vainer and Il'in}(1981{\natexlab{b}})}]{vainer:81a}
\bibinfo{author}{\bibfnamefont{V.~S.} \bibnamefont{Vainer}} \bibnamefont{and}
  \bibinfo{author}{\bibfnamefont{V.~A.} \bibnamefont{Il'in}},
  \bibinfo{journal}{Sov. Phys. Solid State} \textbf{\bibinfo{volume}{23}},
  \bibinfo{pages}{1432} (\bibinfo{year}{1981}{\natexlab{b}}).

\bibitem[{\citenamefont{Kawasuso}()}]{kawasuso:privcom}
\bibinfo{author}{\bibfnamefont{A.}~\bibnamefont{Kawasuso}}, \bibinfo{note}{priv.
  commun.}

\bibitem[{\citenamefont{Puska and Nieminen}(1994)}]{puska:94}
\bibinfo{author}{\bibfnamefont{M.~J.} \bibnamefont{Puska}} \bibnamefont{and}
  \bibinfo{author}{\bibfnamefont{R.~M.} \bibnamefont{Nieminen}},
  \bibinfo{journal}{Rev.~Mod.~Phys.} \textbf{\bibinfo{volume}{66}},
  \bibinfo{pages}{841} (\bibinfo{year}{1994}).

\bibitem[{\citenamefont{Aradi et~al.}(2001)\citenamefont{Aradi, Gali,
  De{\'{a}}k, Lowther, Son, Jan{\'{z}}en, and Choyke}}]{aradi:01}
\bibinfo{author}{\bibfnamefont{B.}~\bibnamefont{Aradi}},
  \bibinfo{author}{\bibfnamefont{A.}~\bibnamefont{Gali}},
  \bibinfo{author}{\bibfnamefont{P.}~\bibnamefont{De{\'{a}}k}},
  \bibinfo{author}{\bibfnamefont{J.~E.} \bibnamefont{Lowther}},
  \bibinfo{author}{\bibfnamefont{N.~T.} \bibnamefont{Son}},
  \bibinfo{author}{\bibfnamefont{E.}~\bibnamefont{Jan{\'{z}}en}},
  \bibnamefont{and} \bibinfo{author}{\bibfnamefont{W.~J.}
  \bibnamefont{Choyke}}, \bibinfo{journal}{Phys.~Rev.~B}
  \textbf{\bibinfo{volume}{63}}, \bibinfo{pages}{245202}
  (\bibinfo{year}{2001}).

\bibitem[{\citenamefont{Gali et~al.}(2000)\citenamefont{Gali, Aradi,
  De{\'{a}}k, Choyke, and Son}}]{gali:00}
\bibinfo{author}{\bibfnamefont{A.}~\bibnamefont{Gali}},
  \bibinfo{author}{\bibfnamefont{B.}~\bibnamefont{Aradi}},
  \bibinfo{author}{\bibfnamefont{P.}~\bibnamefont{De{\'{a}}k}},
  \bibinfo{author}{\bibfnamefont{W.~J.} \bibnamefont{Choyke}},
  \bibnamefont{and} \bibinfo{author}{\bibfnamefont{N.~T.} \bibnamefont{Son}},
  \bibinfo{journal}{Phys.~Rev.~Lett.} \textbf{\bibinfo{volume}{84}},
  \bibinfo{pages}{4926} (\bibinfo{year}{2000}).

\bibitem[{\citenamefont{Petrenko et~al.}(2002)\citenamefont{Petrenko, Petrenko,
  and Bratus}}]{petrenko:02}
\bibinfo{author}{\bibfnamefont{T.~T.} \bibnamefont{Petrenko}},
  \bibinfo{author}{\bibfnamefont{T.~L.} \bibnamefont{Petrenko}},
  \bibnamefont{and} \bibinfo{author}{\bibfnamefont{V.~Y.}
  \bibnamefont{Bratus}}, \bibinfo{journal}{J.~Phys.~Cond.~Matt.}
  \textbf{\bibinfo{volume}{14}}, \bibinfo{pages}{12433} (\bibinfo{year}{2002}).

\bibitem[{\citenamefont{Gali et~al.}(2003{\natexlab{a}})\citenamefont{Gali,
  De{\'{a}}k, Son, Jan{\'{z}}en, von Bardeleben, and Monge}}]{gali:03}
\bibinfo{author}{\bibfnamefont{A.}~\bibnamefont{Gali}},
  \bibinfo{author}{\bibfnamefont{P.}~\bibnamefont{De{\'{a}}k}},
  \bibinfo{author}{\bibfnamefont{N.~T.} \bibnamefont{Son}},
  \bibinfo{author}{\bibfnamefont{E.}~\bibnamefont{Jan{\'{z}}en}},
  \bibinfo{author}{\bibfnamefont{H.~J.} \bibnamefont{von Bardeleben}},
  \bibnamefont{and} \bibinfo{author}{\bibfnamefont{J.-L.} \bibnamefont{Monge}},
  \bibinfo{journal}{Mater.~Sci.~Forum} \textbf{\bibinfo{volume}{433-436}},
  \bibinfo{pages}{511} (\bibinfo{year}{2003}{\natexlab{a}}).

\bibitem[{\citenamefont{Eberlein et~al.}(2002)\citenamefont{Eberlein, Fall,
  Jones, Briddon, and {\"{O}}berg}}]{eberlein:02}
\bibinfo{author}{\bibfnamefont{T.~A.~G.} \bibnamefont{Eberlein}},
  \bibinfo{author}{\bibfnamefont{C.~J.} \bibnamefont{Fall}},
  \bibinfo{author}{\bibfnamefont{R.}~\bibnamefont{Jones}},
  \bibinfo{author}{\bibfnamefont{P.~R.} \bibnamefont{Briddon}},
  \bibnamefont{and}
  \bibinfo{author}{\bibfnamefont{S.}~\bibnamefont{{\"{O}}berg}},
  \bibinfo{journal}{Phys.~Rev.~B} \textbf{\bibinfo{volume}{65}},
  \bibinfo{pages}{184108} (\bibinfo{year}{2002}).

\bibitem[{\citenamefont{Gali et~al.}(2003{\natexlab{b}})\citenamefont{Gali,
  De{\'{a}}k, Rauls, Son, Ivanov, Carlsson, Janzen, and Choyke}}]{gali:03a}
\bibinfo{author}{\bibfnamefont{A.}~\bibnamefont{Gali}},
  \bibinfo{author}{\bibfnamefont{P.}~\bibnamefont{De{\'{a}}k}},
  \bibinfo{author}{\bibfnamefont{E.}~\bibnamefont{Rauls}},
  \bibinfo{author}{\bibfnamefont{N.~T.} \bibnamefont{Son}},
  \bibinfo{author}{\bibfnamefont{I.~G.} \bibnamefont{Ivanov}},
  \bibinfo{author}{\bibfnamefont{F.~H.~C.} \bibnamefont{Carlsson}},
  \bibinfo{author}{\bibfnamefont{E.}~\bibnamefont{Janz{\'{e}}n}}, \bibnamefont{and}
  \bibinfo{author}{\bibfnamefont{W.~J.} \bibnamefont{Choyke}},
  \bibinfo{journal}{Phys.~Rev.~B} \textbf{\bibinfo{volume}{67}},
  \bibinfo{pages}{155203} (\bibinfo{year}{2003}{\natexlab{b}}).

\bibitem[{\citenamefont{Mattausch
  et~al.}(2001{\natexlab{b}})\citenamefont{Mattausch, Bockstedte, and
  Pankratov}}]{mattausch:02}
\bibinfo{author}{\bibfnamefont{A.}~\bibnamefont{Mattausch}},
  \bibinfo{author}{\bibfnamefont{M.}~\bibnamefont{Bockstedte}},
  \bibnamefont{and}
  \bibinfo{author}{\bibfnamefont{O.}~\bibnamefont{Pankratov}},
  \bibinfo{journal}{Physica B} \textbf{\bibinfo{volume}{308-310}},
  \bibinfo{pages}{656} (\bibinfo{year}{2001}{\natexlab{b}}).

\bibitem[{\citenamefont{Eaglesham et~al.}(1994)\citenamefont{Eaglesham, Stolk,
  Gossmann, and Poate}}]{eaglesham:94}
\bibinfo{author}{\bibfnamefont{D.~J.} \bibnamefont{Eaglesham}},
  \bibinfo{author}{\bibfnamefont{P.~A.} \bibnamefont{Stolk}},
  \bibinfo{author}{\bibfnamefont{H.-J.} \bibnamefont{Gossmann}},
  \bibnamefont{and} \bibinfo{author}{\bibfnamefont{J.~M.} \bibnamefont{Poate}},
  \bibinfo{journal}{Appl.~Phys.~Lett.} \textbf{\bibinfo{volume}{65}},
  \bibinfo{pages}{2305} (\bibinfo{year}{1994}).

\bibitem[{\citenamefont{Son et~al.}(2001{\natexlab{b}})\citenamefont{Son, Hai,
  and Janz{\'{e}}n}}]{son:01b}
\bibinfo{author}{\bibfnamefont{N.~T.} \bibnamefont{Son}},
  \bibinfo{author}{\bibfnamefont{P.~N.} \bibnamefont{Hai}}, \bibnamefont{and}
  \bibinfo{author}{\bibfnamefont{E.}~\bibnamefont{Janz{\'{e}}n}},
  \bibinfo{journal}{Mater.~Sci.~Forum} \textbf{\bibinfo{volume}{353-356}},
  \bibinfo{pages}{499} (\bibinfo{year}{2001}{\natexlab{b}}).

\bibitem[{\citenamefont{Zvanut and Konovalov}(2002)}]{zvanut:02}
\bibinfo{author}{\bibfnamefont{M.~E.} \bibnamefont{Zvanut}} \bibnamefont{and}
  \bibinfo{author}{\bibfnamefont{V.~V.} \bibnamefont{Konovalov}},
  \bibinfo{journal}{Appl.~Phys.~Lett.} \textbf{\bibinfo{volume}{80}},
  \bibinfo{pages}{410} (\bibinfo{year}{2002}).

\bibitem[{\citenamefont{Son et~al.}(2002)\citenamefont{Son, Magnusson, and
  Janz{\'e}n}}]{son:02}
\bibinfo{author}{\bibfnamefont{N.~T.} \bibnamefont{Son}},
  \bibinfo{author}{\bibfnamefont{B.}~\bibnamefont{Magnusson}},
  \bibnamefont{and}
  \bibinfo{author}{\bibfnamefont{E.}~\bibnamefont{Janz{\'e}n}},
  \bibinfo{journal}{Appl.~Phys.~Lett.} \textbf{\bibinfo{volume}{81}},
  \bibinfo{pages}{3945} (\bibinfo{year}{2002}).

\end{thebibliography}

\end{document}